\documentclass[manuscript,acmsmall,screen]{acmart}

\usepackage[]{hyperref}
\usepackage{url}
\usepackage{listings}
\usepackage[ruled,vlined]{algorithm2e}
\usepackage{amsmath}

\usepackage{amssymb}
\usepackage{graphicx}
\usepackage{subcaption}
\usepackage{multirow}
\usepackage{pifont}
\usepackage{array}

\definecolor{comment-green}{RGB}{67, 205, 128}
\definecolor{comment-grey}{RGB}{105, 105, 105}

\newcommand{\delr}[1]{}
\newcommand{\addr}[1]{{#1}}

\begin{document}

\setcopyright{acmlicensed}
\acmJournal{TECS}
\acmYear{2025} \acmVolume{1} \acmNumber{1} \acmArticle{1} \acmMonth{1}\acmDOI{10.1145/3760386}

\title{{Re-thinking Memory-Bound Limitations in CGRAs}}

\author{Xiangfeng Liu}
\thanks{We appreciate the reviewers for their insightful and helpful feedback. This work is supported by the National Key Research and Development Program (Grant No. 2024YFB4405600), the National Natural Science Foundation of China (Grant No. 62472086), the Basic Research Program of Jiangsu (Grant No. BK20243042), the Science and Technology Major Special Program of Jiangsu (Grants No. BG2024010), the Start-up Research Fund of Southeast University (Grant No. RF1028624005), and the Provincial Science and Technology Research Project (Grant No. 2024JH2/102400070 and No. 2023JH2/101700370).}
\orcid{0009-0004-7366-5705}
\email{xiangfengliu.cn@gmail.com}
\affiliation{%
  \institution{Northeastern University}
  \city{Shenyang}
  \country{China}
}
\affiliation{%
  \institution{City University of Hong Kong}
  \city{Hong Kong SAR}
  \country{China}
}

\author{Zhe Jiang}
\authornote{Corresponding Author: zhejiang.uk@gmail.com}
\email{zhejiang.uk@gmail.com}
\orcid{0000-0002-8509-3167}
\affiliation{%
  \institution{Southeast University}
  \city{Nanjing}
  \country{China}
}

\author{Anzhen Zhu}
\email{j6434858@gmail.com}
\orcid{0009-0004-7395-6596}
\affiliation{%
  \institution{Northeastern University}
  \city{Shenyang}
  \country{China}
}

\author{Xiaomeng Han}
\email{mingzhihan7@gmail.com}
\orcid{0009-0002-8480-2491}
\affiliation{%
  \institution{Southeast University}
  \city{Nanjing}
  \country{China}
}

\author{Mingsong Lyu}
\email{mingsong.lyu@polyu.edu.hk}
\orcid{0000-0002-4489-745X}
\affiliation{%
  \institution{The Hong Kong Polytechnic University}
  \city{Hong Kong SAR}
  \country{China}
}

\author{Qingxu Deng}
\email{dengqx@mail.neu.edu.cn}
\orcid{0000-0002-5185-6306}
\affiliation{%
  \institution{Northeastern University}
  \city{Shenyang}
  \country{China}
}

\author{Nan Guan}
\email{nanguan@cityu.edu.hk}
\orcid{0000-0003-3775-911X}
\affiliation{%
  \institution{City University of Hong Kong}
  \city{Hong Kong SAR}
  \country{China}
}

\begin{abstract} 

\delr{\textit{Coarse-Grained Reconfigurable Arrays (CGRAs)} are accelerators commonly used to speed up computational workloads that possess iterative structure.}
\addr{\textit{Coarse-Grained Reconfigurable Arrays (CGRAs)} are specialized accelerators commonly employed to boost performance in workloads with iterative structures.}
Existing research typically focuses on compiler or architecture optimizations aimed at improving CGRA performance, energy efficiency, flexibility, and area utilization, under the idealistic assumption that kernels can access all data from Scratchpad Memory (SPM).
\delr{However, when we attempt to use the CGRAs in a real system to accelerate the complicated kernels, such as graph analytics, database operations, and high-performance computing, we encounter significant challenges due to the irregular memory access patterns, reducing the CGRA's utilization of less than 1.5\% and making the CGRA become memory-bound.
To overcome this limitation, we first analyze the root cause of this bottleneck, then redesign the CGRA's memory subsystem and refine the memory model.}
\addr{However, certain complex workloads—particularly in fields like graph analytics, irregular database operations, and specialized forms of high-performance computing (e.g., unstructured mesh simulations)—exhibit irregular memory access patterns that hinder CGRA utilization, sometimes dropping below 1.5\%, making the CGRA memory-bound.
To address this challenge, we conduct a thorough analysis of the underlying causes of performance degradation, then propose a redesigned memory subsystem and refine the memory model.}
With both microarchitectural and theoretical optimization, our solution can effectively manage irregular memory accesses through CGRA-specific runahead execution mechanism and cache reconfiguration techniques. 
Our results demonstrate that we can achieve performance comparable to the original SPM-only system while requiring only 1.27\% of the storage \delr{area}\addr{size}. 
The runahead execution mechanism achieves an average 3.04$\times$ speedup (up to 6.91$\times$), with cache reconfiguration technique providing an additional 6.02\% improvement, significantly enhancing CGRA performance for irregular memory access patterns.

\end{abstract}



\maketitle 

\section{Introduction}
\label{sec:introduction}

\begin{figure}[tbp]
    \centering
    \begin{minipage}{0.48\linewidth}
        \centering
        \includegraphics[width=\linewidth]{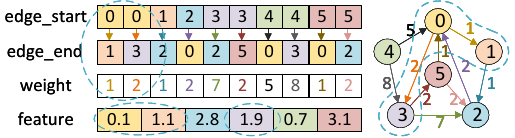}
        \caption{A graph and its edge list, illustrating the start and end nodes with corresponding edge weights. The \textit{feature} array stores the feature values for each node. The dashed circle in the image highlights the graph's edges, nodes, and their edge list entries.}
        \vspace{-6pt}
        \label{fig:graph}
    \end{minipage}
    \hfill
    \begin{minipage}{0.48\linewidth}
        \centering
        \includegraphics[width=\linewidth]{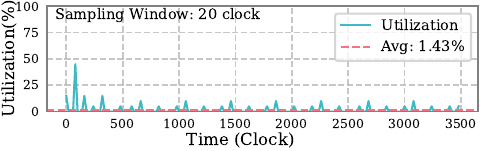}
        \vspace{-21pt}
        \caption{The absence of data in the SPM causes a significant reduction of CGRA utilization, averaging only 1.43\%. Configuration: $4 \times 4$ HyCUBE with 4K SPM~\cite{HyCUBE}. Kernel: Feature aggregation in Graph Convolutional Network using the Cora dataset~\cite{database}.}
        \vspace{-6pt}
        \label{fig:low_utilization}
    \end{minipage}
\end{figure}

\begin{lstlisting}[language=C, caption={Feature aggregation loop in a GCN. This loop iterates over all edges, applying weights to the features of the destination nodes and aggregating the results into the output data for the source nodes.\vspace{-15pt}}, label={lst:feature_aggregation}, float=tbp]
for (i = 0; i < X; i++) {
    output_data[edge_start[i]] += weight[i] * feature[edge_end[i]];}
\end{lstlisting}

\begin{figure*}[tbp]
    \centering
    \includegraphics[width=\linewidth]{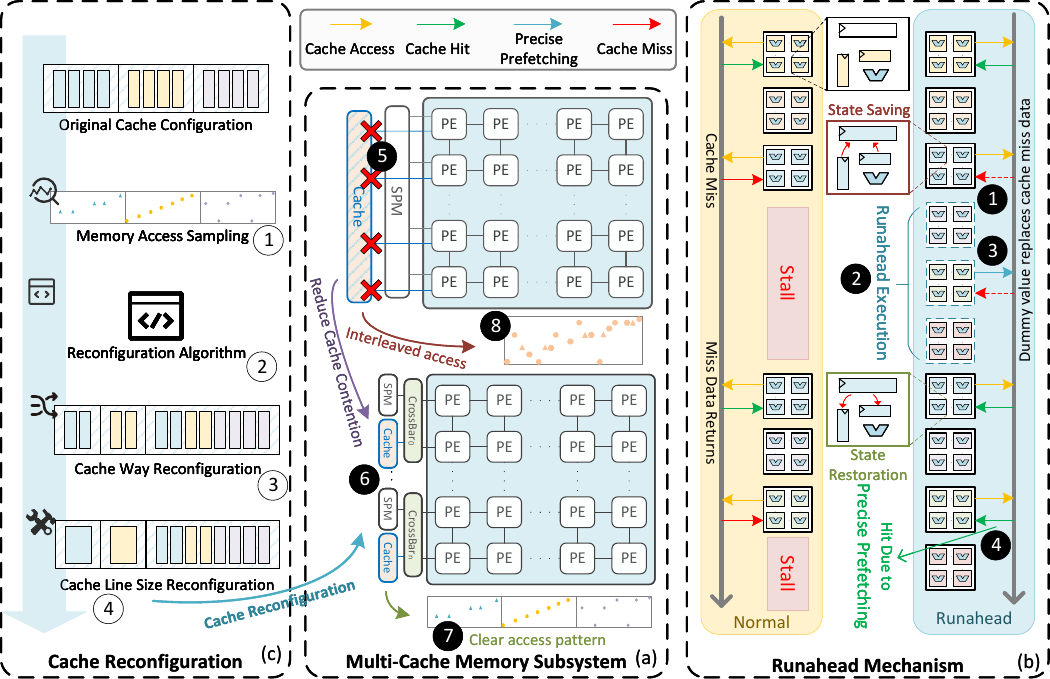}
    \caption{Overview of the Proposed Architecture: (a) Multi-Cache Memory Subsystem: This design illustrates the evolution of our memory subsystem, which reduces cache contention among PEs and enhances scalability through a multi-cache architecture. (b) Runahead Mechanism: By utilizing CGRA stall times for prefetching, this mechanism reduces cache miss rates and improves CGRA performance. (c) Cache Reconfiguration Process: Cache resources are reallocated and customized according to the specific needs of each PE, optimizing resource utilization and further enhancing overall CGRA performance.}
    \label{fig:overview}
\end{figure*}

Recent advancements in science, particularly in AI technologies, increase the demand for accelerated computations. CGRAs consist of configurable networks interconnecting multiple Processing Elements (PEs), each performing computations based on its configuration. \addr{Typically, the memory-access PEs can read from or write to a small, low-latency on-chip buffer called SPM. Under ideal conditions, computation-relevant data is fetched from main memory into the SPM before it is used, reducing data-transfer delays and helping sustain high performance by keeping frequently accessed data close to the PEs. Consequently, reliance on high-latency main memory is alleviated.} This architecture balances power, performance, and flexibility.

CGRAs accelerate computation-intensive portions of programs (e.g., the one shown in Listing~\ref{lst:feature_aggregation}), known as kernels, by compiling them into Data Flow Graphs (DFGs) and mapping nodes to PEs for parallel execution. The network facilitates data transfer, ensuring throughput and correctness.

Previous research primarily focuses on designing compilers for efficient kernel mapping and developing microarchitectures that enhance CGRA performance~\cite{7167295,ChordMap,Ccf,OpenCGRA,Pillars,TAEM,9075353,Morpher,DRIPS,ML_CGRA,Rewriting_History,E2EMap,6861574,CGRA_ME,HyCUBE,Snafu,9407079,Revamp}. Most studies assume an idealized memory model where kernel data fits entirely within the SPM, simplifying memory access management. However, irregular and large-volume memory accesses can significantly degrade performance, leading to challenges in real-world deployments.

For workloads, like Graph Neural Networks (GNNs)~\cite{gnn}, high-performance computing~\cite{hpc}, and databases~\cite{hash}, kernels often process large computational data exceeding SPM capacity. These kernels exhibit irregular memory access patterns due to data structure irregularities or algorithmic characteristics, making future accesses unpredictable and prefetching techniques less effective~\cite{reuse,Static}.
For example, in GNNs, the irregular connectivity of graph structures results in non-contiguous and unpredictable memory accesses (Figure~\ref{fig:graph}). Arrays accessed depend on values stored in other arrays, leading to indirect access patterns that complicate prefetching(Listing~\ref{lst:feature_aggregation})~\cite{DRIPS}.

Under these conditions, our experiments demonstrate that CGRAs encounter stalls when data is absent from the SPM, resulting in utilization dropping as low as 1.43\% (Figure~\ref{fig:low_utilization}). Addressing these memory-related bottlenecks is essential for maximizing the acceleration potential of CGRAs.
To tackle these challenges, we introduce a comprehensive solution, as illustrated in Figure~\ref{fig:overview}. We redesign the CGRA's memory subsystem by incorporating a cache (Figure~\hyperref[fig:overview]{\ref{fig:overview}a~\textcolor{black}{\ding{186}}}). However, irregular accesses lead to poor locality and high cache miss rates. To mitigate this, we introduce a CGRA-specific runahead execution mechanism (Figure~\hyperref[fig:overview]{\ref{fig:overview}b}). Upon a cache miss (\hyperref[fig:overview]{\textcolor{black}{\ding{182}}}), the CGRA saves its state and enters a Runahead state (\hyperref[fig:overview]{\textcolor{black}{\ding{183}}}), where valid memory requests trigger precise prefetching (\hyperref[fig:overview]{\textcolor{black}{\ding{184}}}), loading future required data into the cache (\hyperref[fig:overview]{\textcolor{black}{\ding{185}}}). This approach enhances cache hit rates, achieving an average speedup of $3.04\times$ (up to $6.91\times$). To our knowledge, our work is the first in the CGRA domain to explicitly tackle performance degradation from irregular memory access.

As CGRAs scale, multiple PEs sharing a single cache cause severe contention (\hyperref[fig:overview]{\textcolor{black}{\ding{186}}}). We propose a multi-cache design that scales the number of caches proportionally (\hyperref[fig:overview]{\textcolor{black}{\ding{187}}}), effectively isolating cross-PE conflicts.
In this multi-cache subsystem, PEs have varying demands for cache configurations. We introduce a hardware/software co-designed cache reconfiguration technique that adjusts cache parameters based on individual PE memory access characteristics (Figure~\hyperref[fig:overview]{\ref{fig:overview}c~\textcolor{black}{\ding{172}\ding{188}}}). This involves reallocating cache ways from PEs with lower demands to those with higher demands (\hyperref[fig:overview]{\textcolor{black}{\ding{174}}}) and fine-tuning cache line sizes to align with access patterns (\hyperref[fig:overview]{\textcolor{black}{\ding{175}}}).
The contributions include:

\begin{itemize}
    \item \textbf{Redesigned CGRA's Memory Subsystem with Cache Integration}: We address the limitations of traditional SPM-based CGRAs by integrating caches into the memory subsystem, allowing for more efficient handling of large computational data and irregular memory accesses that exceed SPM capacity.
    
    \item \textbf{CGRA-Specific Runahead Execution Mechanism}: We implement a runahead mechanism for CGRAs, enabling execution during cache misses by substituting dummy values. This significantly improves performance through precise prefetching.
    
    \item \textbf{Multi-Cache Memory Subsystem}: We develop a highly scalable multi-cache design that dynamically increases cache availability as the CGRA architecture scales, effectively reducing contention in larger architectures and ensuring consistent performance gains.

    \item \textbf{Memory-access-pattern-aware cache reconfiguration}: We introduce cache reconfiguration strategies based on PE memory access patterns. Reallocating cache resources and configuring cache parameters according to PE demands enhances overall performance.

\end{itemize}

\section{Preliminaries}
\subsection{Microarchitecture of CGRAs}

\begin{figure}[t]
    \centering
    \includegraphics[width=\linewidth]{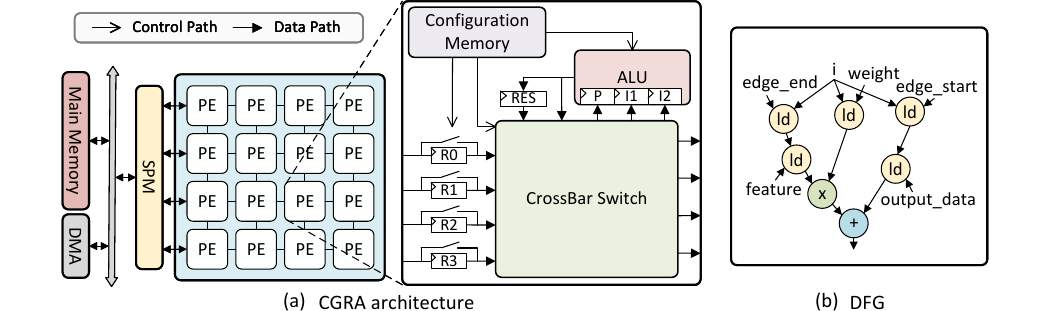}
    \caption{\addr{Overall CGRA architecture and the DFG for listing~\ref{lst:feature_aggregation}. The CGRA comprises an array of PEs interconnected by a crossbar-based configurable network, with edge PEs connected to an SPM. The SPM communicates with main memory through a bus, and data can also be transferred via DMA (Direct Memory Access). Each PE integrates an ALU, crossbar switch, register file, and configuration memory. To accelerate a kernel, it is transformed into a DFG and then mapped onto the PEs by the mapper.}}
    \vspace{-15pt}
    \label{fig:cgra_architecture}
\end{figure}

Typical CGRA architecture (Figure~\hyperref[fig:cgra_architecture]{\ref{fig:cgra_architecture}a})~\cite{HyCUBE,ADRES,CGRA_ME,MorphoSys} consists of an array of PEs interconnected via a \addr{crossbar-base} configurable network. Edge PEs are connected to \delr{a}\addr{an} SPM to facilitate load and store operations. Each PE includes an ALU \addr{(Arithmetic Logic Unit)}, a crossbar switch, a register file, and a configuration memory (\textit{config mem}) that controls the behavior of the crossbar switch and ALU on a cycle basis.

To accelerate a kernel, it is transformed into a DFG (Figure~\hyperref[fig:cgra_architecture]{\ref{fig:cgra_architecture}b} shows the DFG for listing~\ref{lst:feature_aggregation}), and a mapper assigns computation nodes to the PEs. This mapping process generates control signals for the ALUs and crossbar switches within the PEs, which are stored in the \textit{config mem}. These control signals ensure that PEs execute operations and manage data transfers according to the DFG.

\subsection{\addr{Motivation:} Memory-Bound Limitations in CGRA}
\label{sec:Motivation}

In CGRAs, PEs operate deterministically without handshake signals; all PEs must adhere to the control flow specified in the \textit{config mem} to maintain computational correctness. If a memory-accessing PE cannot retrieve the required data, it causes the entire CGRA to stall. Typically, CGRAs store kernel data in the SPM to ensure that memory-accessing PEs can obtain data as expected. However, in many real-world kernels, the data required for computation exceeds the capacity of the SPM, and prefetching strategies are effective only for regular memory access patterns.

\begin{figure}[t]
    \centering
    \includegraphics[width=\linewidth]{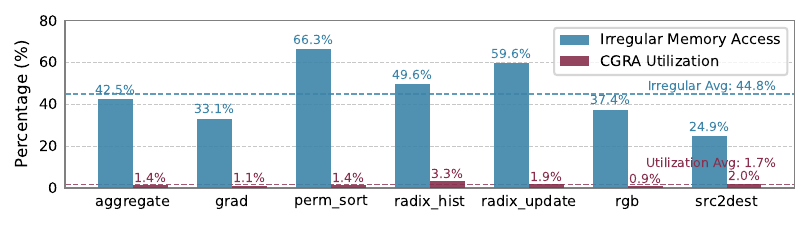}
    \vspace{-28pt}
    \caption{\addr{Proportion of irregular memory access among all memory accesses for various workloads (Table~\ref{tab:benchmarks}), alongside the corresponding CGRA utilization. The findings show that irregular access significantly reduces CGRA utilization to an average of merely 1.7\%, thereby having a substantial impact on overall performance.}}
    \vspace{-20pt}
    \label{fig:irregular_vs_utilization}
\end{figure}

With irregular memory accesses, necessary data may not reside in the SPM when required, forcing PEs to access main memory. This access can take tens of cycles or more, during which the entire CGRA remains stalled, leading to significant performance degradation due to memory-bound limitations. As illustrated in Figure~\ref{fig:low_utilization}, this issue reduces the average CGRA utilization to as low as 1.43\%. \addr{Furthermore, Figure \ref{fig:irregular_vs_utilization} quantifies the proportion of irregular memory accesses across various workloads (refer to  Table \ref{tab:benchmarks}) and correlates them with the corresponding CGRA utilization. The results reveal that irregular accesses can reduce the CGRA utilization to an average of just 1.7\%, substantially affecting overall performance.} Such challenges are common across CGRAs and limit their broader applicability\addr{, underscoring the urgent need for architectural optimizations that efficiently handle irregular memory accesses}.

To address these limitations, we develop a new memory subsystem for CGRAs designed to mitigate performance degradation caused by irregular and large-volume memory accesses. We implement and evaluate these methods within our simulation framework, demonstrating their effectiveness in enhancing CGRA performance.

Our proposed techniques are broadly applicable to various types of CGRAs. The runahead execution mechanism requires only the addition of state-switching functionality and dummy data tracking capabilities to the base PEs, given that a CGRA comprises multiple instantiated PEs. The cache reconfiguration technique does not necessitate any changes to the overall CGRA architecture.

\section{Mitigating Memory-Bound Limitations}

This section presents a comprehensive methodology to address the multiple memory-related challenges inherent in CGRAs. Our solution comprises four key components, each targeting specific issues with innovative strategies. In Section~\ref{sec:cache_integrate}, we redesign the memory subsystem by integrating a cache hierarchy, enabling CGRAs to handle computational data exceeding the capacity of the SPM and to manage irregular memory accesses effectively (Figure~\hyperref[fig:overview]{\ref{fig:overview}a~\textcolor{black}{\ding{186}}}). However, our simulations reveal that irregular memory access patterns significantly reduce memory locality, leading to increased cache miss rates and frequent CGRA stalls. To mitigate this issue, Section~\ref{sec:cgra_runahead} introduces a CGRA-specific runahead execution mechanism (Figure~\hyperref[fig:overview]{\ref{fig:overview}b}). When a cache miss occurs, the CGRA no longer passively stalls; instead, it proactively utilizes the stall time for data prefetching, significantly enhancing cache hit rates and improving overall CGRA performance. As CGRAs scale up, severe contention from sharing a single cache among multiple PEs emerges as a new challenge. Section~\ref{sec:multi-cache} addresses this by proposing a multi-cache architecture that scales the number of caches proportionally with the size of the CGRA, effectively alleviating cache contention in large-scale systems (Figure~\hyperref[fig:overview]{\ref{fig:overview}a~\textcolor{black}{\ding{187}}}). Finally, recognizing the unique memory access characteristics and customized cache requirements of different PEs, Section~\ref{sec:cache_reconfig} develops a cache reconfiguration technique (Figure~\hyperref[fig:overview]{\ref{fig:overview}c}). This dynamic approach reallocates cache resources by shifting cache ways from low-demand to high-demand PEs and adjusts cache line sizes based on the specific memory access characteristics of PEs. This multifaceted approach not only resolves CGRA data capacity limitations but also significantly enhances their ability to process kernels with irregular memory access patterns while ensuring that the memory subsystem scales efficiently with CGRA growth.

\subsection{Cache-Integrated Memory Subsystem}
\label{sec:cache_integrate}

\begin{figure}[t]
    \centering
    \includegraphics[width=\linewidth]{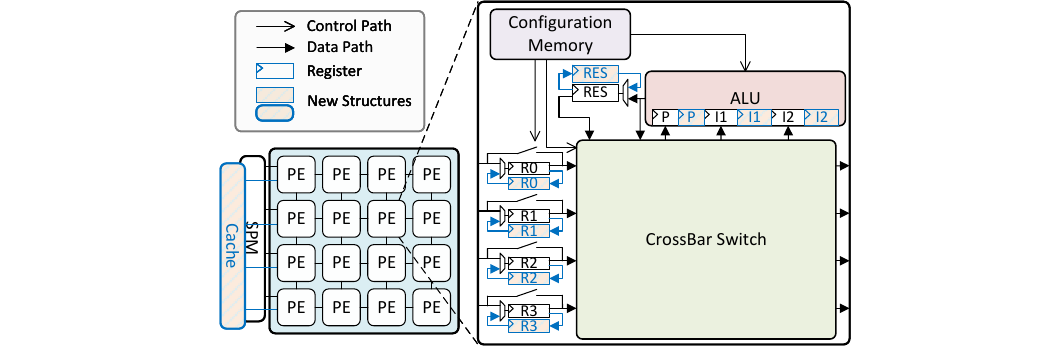}
    \vspace{-20pt}
    \caption{Integration of cache mechanisms into the CGRA and addition of state save and restore logic.}
    \vspace{-10pt}
    \label{fig:runahead_logic}
\end{figure}

As depicted in Figure~\ref{fig:runahead_logic}, we develop a new memory subsystem for the CGRA that integrates \delr{a}\addr{an} SPM with a cache hierarchy comprising L1 and L2 caches. The SPM stores a portion of the computational data, while data exceeding its capacity is accessed through the cache hierarchy. The L1 cache is optimized for low-latency access, and the non-inclusive L2 cache is designed to achieve a higher hit rate. We select this hybrid approach over simply expanding the SPM or entirely replacing it with caches for several reasons:

\begin{itemize}
\item \textbf{Scalability and Energy Efficiency}: Expanding the SPM to accommodate workloads with irregular memory access patterns results in significant area and power overheads. Caches enhance scalability and energy efficiency, especially for these workloads.
\item \textbf{Flexibility and Adaptability}: While SPMs are highly efficient for predictable memory access, caches excel at managing irregular accesses. Combining both allows the CGRA to effectively process a broader range of kernels with diverse memory access characteristics.
\item \textbf{Programming Simplicity}: Incorporating caches reduces the burden on programmers to manually manage data movement, which would be necessary when using SPMs alone.
\end{itemize}

When the CGRA issues memory access requests, a crossbar switch routes them to either the SPM or the cache. In a single cycle, the CGRA may generate multiple memory access requests; these are arbitrated and sent sequentially to the L1 cache for processing. This design enables the CGRA to handle kernels with data exceeding the SPM capacity and to retrieve data for kernels with irregular memory access patterns via the cache hierarchy.

\subsection{Runahead Execution in CGRAs}
\label{sec:cgra_runahead}

While our memory subsystem addresses the challenges of handling kernels with data exceeding the capacity of the SPM and irregular memory access patterns, cache misses can still cause the CGRA to stall while waiting for data retrieval, significantly impacting performance. To better utilize stall time and enhance CGRA efficiency, we draw inspiration from the concept of \textit{runahead execution} in processor design~\cite{runahead1}. However, it is important to note that our implementation diverges significantly from traditional processor-based runahead techniques. Unlike CPUs—which have complex control logic, deep pipelines, and speculative execution capabilities—CGRAs represent a fundamentally different architectural paradigm with distributed arrays of PEs that operate with simpler control and deterministic execution.

Our runahead mechanism enables precise prefetching of future data during stall times caused by cache misses. As illustrated in Figure~\hyperref[fig:overview]{\ref{fig:overview}b~\textcolor{black}{\ding{182}}}, when the CGRA encounters a stall due to a cache miss, it saves its current state and transitions into a runahead state (\hyperref[fig:overview]{\textcolor{black}{\ding{183}}}), instead of pausing completely as in normal execution (shown on the left). In this runahead state, the CGRA substitutes missing data with dummy values and continues execution. Concurrently, valid memory access requests trigger precise prefetching (\hyperref[fig:overview]{\textcolor{black}{\ding{184}}}), allowing the CGRA to load future required data into the cache as quickly as possible. The system meticulously tracks the propagation of these dummy values; any invalid read or write operations—where either the address or data depends on dummy values—are discarded. To prevent state corruption upon resuming normal execution, valid write operations are redirected to a temporary storage area instead of the cache or SPM.

During runahead execution, write operations are converted into corresponding read operations to trigger prefetching, ensuring no irreversible impact on the system state. Specifically, these writes are never actually committed to memory and serve solely for prefetching purposes, preserving the integrity of the CGRA system state and eliminating the need for rollback upon runahead completion. If invalid data reads occur—where the memory address depends on previously used dummy values—dummy values are used directly. For valid data reads, where the address does not depend on dummy values, the request is sent to the memory subsystem or temporary storage. If the request hits in the SPM, cache, or temporary storage, the retrieved data is used to continue execution. In the case of a cache miss, dummy values are used to allow execution to proceed, while the cache prefetches the relevant data. When the original cache miss that triggered the runahead process is resolved (multiple cache misses may occur, as multiple memory access requests can be issued in a single CGRA cycle), the CGRA restores its previous state and returns to normal execution. Since the necessary data will have already been prefetched during the runahead execution, subsequent cache misses are avoided (\hyperref[fig:overview]{\textcolor{black}{\ding{185}}}), preventing further stalls as would occur in normal execution(shown on the left). This approach significantly enhances cache hit rates and reduces stalls.

However, due to the influence of dummy values, not all valid memory requests issued during runahead can be guaranteed to be useful in future execution. For instance, if a memory write operation is incorrectly updated with a dummy value and that data is later read and used, the execution may enter an incorrect state. For example, if a memory write operation does not update data at an address because of dummy values, and this data is subsequently read and utilized, the execution may become inconsistent.

\subsubsection{Key Design Aspects of CGRA Runahead Execution}

Our CGRA runahead execution mechanism incorporates several critical design elements to enhance performance and ensure data integrity:

\begin{itemize}
\item \textbf{State Management}: We implement a state save and restore mechanism using dedicated backup registers (Figure~\ref{fig:runahead_logic}). This design enables seamless transitions between normal and runahead execution, effectively leveraging idle cycles during cache misses.
\item \textbf{Dummy Data Handling and Selective Prefetching}: We introduce a dummy data tracking system using additional flag bits and modify existing ALUs to track dummy data propagation. This ensures high-quality, targeted prefetching, significantly reducing cache pollution and optimizing cache utilization.
\item \textbf{Temporary Storage Strategy}: To maintain data consistency without discarding potentially useful information, we implement a temporary storage area. After evaluating two approaches, we found that partitioning the SPM is more effective than repurposing existing cache space, as it minimizes cache disruption and provides dedicated space for runahead execution.
\item \textbf{Cache Policy Optimization}: We configure the cache as non-blocking with a Least Recently Used (LRU) replacement policy and a write-allocate strategy. These choices support runahead execution, efficiently manage prefetched data, and enable prefetching for future operations.
\end{itemize}

Through these considerations, our CGRA runahead mechanism maximizes computational efficiency by utilizing idle cycles during cache misses, maintains data integrity and execution correctness despite the use of runahead execution, and enhances memory subsystem performance while minimizing cache pollution through targeted, high-quality prefetching and sensible cache policy choices. 
By comprehensively addressing these aspects, our runahead mechanism delivers a substantial enhancement in CGRA performance. This improvement is particularly pronounced for workloads that surpass SPM capacity and are characterized by irregular memory access patterns, ensuring efficient execution and optimal resource utilization.

\subsection{Memory Subsystem with Multiple L1 Caches}
\label{sec:multi-cache}

As the CGRA scales up, the number of memory-accessing PEs increases, leading to potential cache contention when multiple memory accesses are triggered within the same cycle. This contention causes frequent stalls in the CGRA, significantly degrading performance. To address this issue and ensure scalability, we propose a multi-cache memory subsystem comprising multiple L1 caches that share a common L2 cache. As the CGRA grows, the number of L1 caches increases proportionally, maintaining consistent scalability between the memory subsystem and the CGRA.

We opt to partition only the L1 cache for several key reasons. First, since L1 caches are closest to the PEs and have the most significant impact on access latency, partitioning them minimizes contention at this critical level. Second, PEs often exhibit different memory access patterns, and multiple L1 caches can better capture these patterns without interference from other PEs. Third, partitioning only the L1 caches strikes a balance between performance improvement and the additional power associated with cache partitioning. This design choice addresses scalability challenges at the L1 level, where contention is severe, while maintaining the benefits of a shared, larger L2 cache for overall system performance.

\addr{\textit{Coherence Across Multiple Caches.}
 Although classical consistency protocols can address coherence issues, leveraging CGRA's unique compilation and architectural features allows us to elegantly and effectively tackle multi-cache coherence in CGRA systems. In our design, each crossbar and its associated physical SPM and cache operate as a single "virtual SPM" (see Figure~\hyperref[fig:overview]{\ref{fig:overview}a~\textcolor{black}{\ding{187}}}). Each crossbar is shared by a small number of border PEs (two in Figure~\hyperref[fig:overview]{\ref{fig:overview}a~\textcolor{black}{\ding{187}}}), ensuring that each PE directly accesses only its assigned virtual SPM. During compilation, the CGRA's compile-time data allocation and static scheduling mechanisms partition data among these virtual SPMs according to the computational needs of the connected PEs, ensuring each PE can access its required data. Because the data is fully partitioned across separate virtual SPMs (with no overlap), each cache only handles data from its own partition, thereby eliminating inter-cache consistency conflicts. Within a single virtual SPM, multiple PEs may share the same cache, but their operations follow a statically determined dataflow schedule, preventing potential conflicts over data consistency. Through these design choices, we effectively address multi-cache coherence in our CGRA system.
}

\subsection{Cache Reconfiguration}
\label{sec:cache_reconfig}

\begin{figure}[t]
    \centering
    \includegraphics[width=\linewidth]{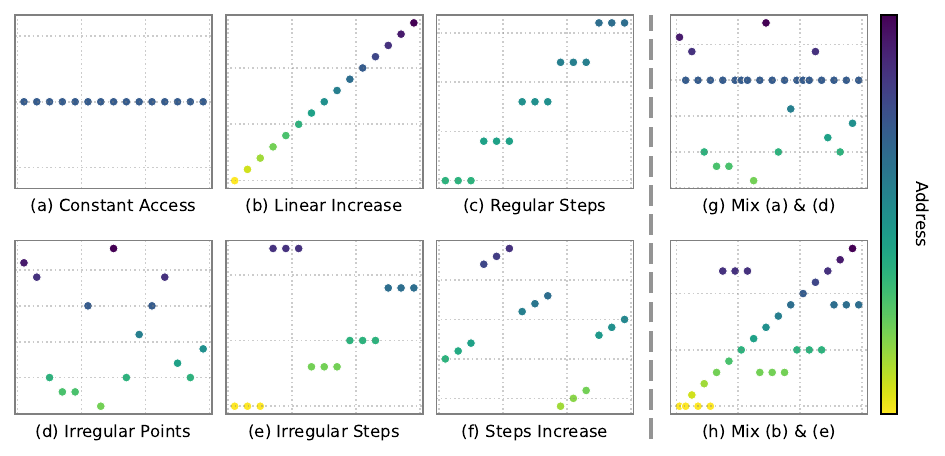}
    \vspace{-20pt}
    \caption{Memory Access Patterns from Benchmark: Top-left shows regular accesses; bottom-left shows irregular accesses; right shows a mix of both (x-$axis$: Time, y-$axis$: Address)}
    \label{fig:memory_patterns}
\end{figure}

In our multi-cache memory subsystem, we observe that memory-accessing PEs exhibit unique and distinct memory access patterns, as illustrated in Figure~\hyperref[fig:overview]{\ref{fig:overview}a~\textcolor{black}{\ding{188}}}. We categorize these patterns, depicted in Figure~\ref{fig:memory_patterns}, into two types: \textbf{Regular Memory Accesses}, which show high predictability with strong correlations in adjacent memory addresses—for example, constant addresses (Figure~\hyperref[fig:memory_patterns]{\ref{fig:memory_patterns}a}), linear increments or decrements (Figure~\hyperref[fig:memory_patterns]{\ref{fig:memory_patterns}b}), and regular steps (Figure~\hyperref[fig:memory_patterns]{\ref{fig:memory_patterns}c}); and \textbf{Irregular Memory Accesses}, which exhibit low predictability where adjacent accesses do not follow a strong pattern—for example, completely random access points (Figure~\hyperref[fig:memory_patterns]{\ref{fig:memory_patterns}d}) and irregular steps (Figures~\hyperref[fig:memory_patterns]{\ref{fig:memory_patterns}e} and \hyperref[fig:memory_patterns]{\ref{fig:memory_patterns}f}). Specific patterns may combine elements of both types.

In \textbf{Regular Memory Accesses}, the strong correlation between adjacent accesses means that data loaded into the cache is likely to be used in subsequent operations. Therefore, larger cache line sizes are beneficial, effectively prefetching data that will soon be needed. Increasing cache size and associativity provides limited benefits, as data is typically used sequentially and not frequently reused. Cache thrashing may occur between different regular access streams, in which case higher associativity can help mitigate this issue.

In contrast, \textbf{Irregular Memory Accesses} lack strong patterns, making it challenging to predict future data needs. Consequently, larger cache sizes are advantageous since they store more data, increasing the likelihood of future cache hits. Higher cache associativity reduces cache thrashing, which is more likely with irregular access patterns compared to regular ones. Cache line size requires careful tuning: too large may lead to prefetching unnecessary data and cause frequent replacements, while too small may not prefetch enough useful data.

These distinct memory access patterns of CGRA memory-accessing PEs are previously obscured when sharing a single cache, as illustrated in Figure~\hyperref[fig:overview]{\ref{fig:overview}a~\textcolor{black}{\ding{189}}}. This phenomenon is analogous to memory accesses in processors where various regular access patterns are interleaved, effectively masking their inherent regularity~\cite{ReCA}.

\begin{figure*}[tbp]
    \centering
    \includegraphics[width=\linewidth]{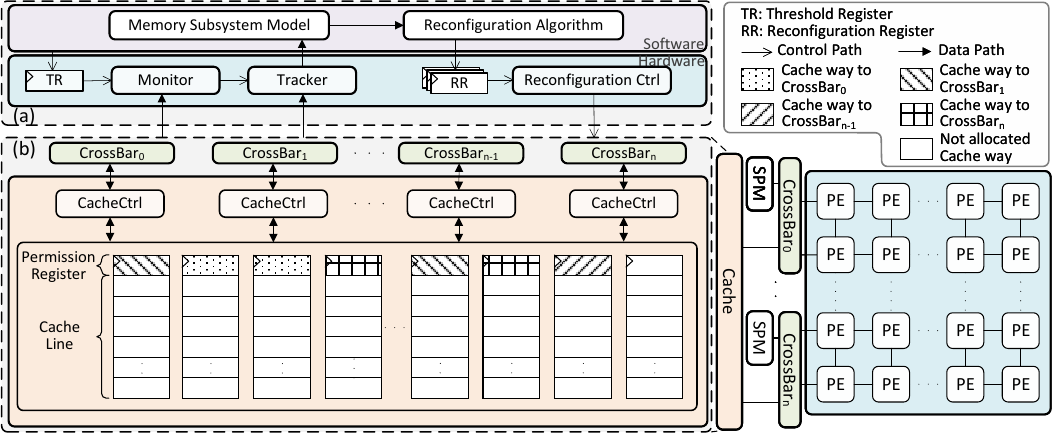}
    \caption{The top-level architecture of the memory subsystem supports cache reconfiguration via hardware monitoring and sampling of memory access information, software analysis and decision-making, and assignment of different cache ways by a reconfiguration controller based on Permission Registers. This enables flexible resource allocation tailored to the demands of individual PEs. \addr{Each crossbar is shared by two border PEs, with n representing the total number of crossbars and their corresponding SPM modules. Each PE pair accesses the appropriate SPM or cache data through its designated crossbar.}}
    \vspace{-10pt}
    \label{fig:top_level_cache_reconfig}
\end{figure*}

The varying cache configuration requirements driven by these diverse memory access patterns, coupled with the variability of CGRA-accelerated kernels, inspire us to propose reconfigurable caches tailored to the memory access patterns exhibited by memory-accessing PEs. As depicted in Figure~\hyperref[fig:top_level_cache_reconfig]{\ref{fig:top_level_cache_reconfig}a}, the hardware architecture integrates MMIO-mapped threshold and reconfiguration registers. A dedicated hardware monitor continuously observes cache miss rates, triggering the hardware tracker when a predefined threshold is exceeded. The tracker then samples each PE's memory accesses over a configurable observation window.

Upon sample completion, the data is delivered to the software via an interrupt-driven mechanism. The software employs a two-phase approach: First, it uses a built-in memory subsystem model to measure the hit rates of individual L1 caches across various resource allocation scenarios, deliberately ignoring total L1 cache resource constraints during this profiling phase. Subsequently, leveraging the collected hit rate data, it applies a reconfiguration algorithm to generate an optimal resource allocation strategy under strict total L1 cache resource limitations. The optimized strategy is then translated into configuration instructions and deployed through MMIO-based APIs to update the reconfiguration registers.

The hardware reconfiguration controller applies new configurations by modifying relevant control registers (e.g., the Cache Way Permission Register) according to the settings stored in the reconfiguration registers. This closed-loop adaptation mechanism enables dynamic redistribution of cache resources based on runtime access patterns, thereby optimizing resource utilization and enhancing overall system performance.

\subsubsection{Reconfigurable Cache Architecture}

The CGRA's non-blocking cache employs two critical data structures—the \textit{Miss Status Handling Register (MSHR)} and the \textit{Load/Store Table}—to store cache miss information and enable the handling of new memory requests during a miss.

\begin{figure}[tbp]
    \centering
    \begin{subfigure}{0.38\linewidth}
        \centering
        \includegraphics[width=\linewidth]{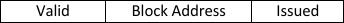}
        \caption{MSHR fields: \textit{Block Address} captures the \textit{block address} for cache misses, and \textit{Issued} indicates whether the data request has been dispatched.}
        \label{fig:mshr}
    \end{subfigure}    
    \hfill
    \begin{subfigure}{0.58\linewidth}
        \centering
        \includegraphics[width=\linewidth]{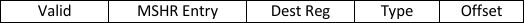}
        \caption{Load/Store Table fields: \textit{MSHR Entry} identifies the associated MSHR, \textit{Dest Reg} indicates the source of the miss request, \textit{Type} specifies whether the operation is a read or a store, \textit{Offset} denotes the position relative to the cache block.}
        \label{fig:load_store_table}
    \end{subfigure}
    \vspace{-10pt}
    \caption{MSHR and Load/Store Table}
    \vspace{-10pt}
    \label{fig:MSHR_Load_store_table}
\end{figure}

Each MSHR entry, illustrated in Figure~\ref{fig:mshr}, consists of:
\begin{itemize}
\item \textbf{Valid}: Marks entry occupancy; clears when the corresponding cache line is retrieved.
\item \textbf{Block Address}: The starting address of the cache line. For instance, with a 32-bit address and a 16-byte cache line, 28 bits represent the \textit{Block Address}.
\item \textbf{Issued}: Denotes whether the associated data request has been dispatched to the next-level memory, as some requests might not be issued immediately.
\end{itemize}

The Load/Store Table, shown in Figure~\ref{fig:load_store_table}, records details of each cache miss. Each entry includes:
\begin{itemize}
\item \textbf{MSHR Entry}: References the corresponding MSHR entry associated with the miss.
\item \textbf{Dest Reg}:
\textit{Read Misses}: Indicates that the miss caused the CGRA to enter runahead mode. When the required data block arrives, it is sent to the CGRA, and normal execution resumes.\textit{Write Misses}: Records the position of the write data in the \textit{Store Buffer}, a dedicated storage unit. Upon data block retrieval, the data is merged with the \textit{Store Buffer} content, written to the cache, and the \textit{Store Buffer} entry is released.
\item \textbf{Type}: Specifies the instruction type corresponding to the miss (e.g., LW for Load Word, LH for Load Half Word, SW for Store Word), used for data extraction when resolving the miss.
\end{itemize}

\paragraph{Cache Size Reconfiguration}

As shown in Figure~\hyperref[fig:top_level_cache_reconfig]{\ref{fig:top_level_cache_reconfig}b}, each cache way is assigned a \textit{permission register} that specifies its associated virtual SPM. By distributing cache ways among different virtual SPMs, we enable reconfiguration of cache size and associativity. Allocating at the cache way level offers two main advantages. First, it adheres to binary address indexing constraints—which require the number of sets to be a power of two—thereby maintaining structural integrity. Second, it simplifies hardware design by allowing cache size adjustments to meet associativity requirements without introducing additional complexity or overhead.

\paragraph{Cache Line Size Reconfiguration}

Cache line size reconfiguration involves merging adjacent physical cache lines within a cache way to form a \textit{virtual cache line}. In this context, a traditional cache line is referred to as a physical cache line, while a virtual cache line comprises \( 2^m \) physical cache lines, where \( m \in \mathbb{Z}_{\geq 0} \)~\cite{self_tuning}.

Cache line replacements occur at the virtual cache line level, involving multiple physical cache lines. However, if a cache access results in a hit, read and write operations continue at the physical cache line granularity. Because the L2 cache line size matches the maximum cache line size of the L1 cache, physical cache lines within a virtual cache line can only result in full hits or full misses, eliminating partial hits. Therefore, treating the virtual cache line as a single unit and processing its physical cache lines sequentially is justified.

For implementing LRU with virtual cache lines, we shift from prioritizing ways within a physical cache set to ways within a virtual cache set. Specifically, the first physical cache set within the virtual cache set serves as the representative. All requests to the virtual cache set, after appropriate masking, are treated as requests to this first physical cache set. When selecting a cache way for LRU replacement, we use the data corresponding to this representative set.

\subsubsection{Cache Reconfiguration Strategy} 

Let \( A \) represent the number of L1 caches, \( S \) be the total cache size(ways), and \( L \) be the set of valid cache line sizes. For \( i = 0, 1, \dots, A-1 \), let \( S_i \) denote the number of cache ways allocated to the \( i \)-th L1 cache, and \( L_i \in L \) represent its cache line size. The function \( h_i(L_i, S_i) \) represents the hit rate of the \( i \)-th cache with a given cache line size \( L_i \) and cache size \( S_i \), while \( H_i(S_i) = \max_{L_i} h_i(L_i, S_i) \) denotes the maximum hit rate for the \( i \)-th cache at cache size \( S_i \).

Our objective is to maximize the sum of the logarithms of \( H_i(S_i) \) for all L1 caches:
\begin{equation}
\max_{\{S_i\}} \quad \sum_{i=0}^{A-1} \log H_i(S_i)
\label{eq:objective}
\end{equation}
Subject to the following constraints:
\begin{gather}
\sum_{i=0}^{A-1} S_i \leq S \label{eq:constraint1} \\
\delr{S_i \geq 0, \quad }S_i \in \mathbb{Z}_{\geq 0}, \quad \text{for} \quad i = 0, 1, \dots, A-1 \label{eq:constraint2}
\end{gather}

The goal is to allocate cache resources optimally, given the total cache size constraint, in a way that maximizes overall performance. Since a single cache miss in \delr{an}\addr{a} CGRA can cause the entire array to stall, optimal performance is achieved when all memory accesses result in cache hits during each cycle. Therefore, the objective is to maximize the product of hit rates across all caches\footnote{This is not a strict mathematical condition but serves as a more effective and practical heuristic for improving cache size allocation in synchronized environments.}. Taking the logarithm of the product converts this into a sum, thereby simplifying the original problem into a linear programming formulation.

To determine the optimal hit rates \( H_i(S_{i}) \) for different cache configurations, we develop a Memory Subsystem Model to simulate cache behavior under different configurations. Using sampled memory access data, we derive the hit rates for each \( i \)-th L1 cache under various cache configurations.

\begin{figure*}[tbp]
  \centering
  \begin{minipage}{0.66\linewidth}
  \begin{algorithm}[H]
\small
\SetAlgoLined
\DontPrintSemicolon
\SetAlgoVlined
\SetKwProg{Function}{Function}{}{}
\SetKwFunction{MaxProfit}{max\_profit}
\SetKwInput{Input}{Input}
\SetKwInput{Output}{Output}
\SetKw{KwTo}{to}

\Function{\MaxProfit{$\mathcal{H}$, $T_{max}$} (Time Complexity: $\mathcal{O}(n \cdot T_{max}^2)$)}{
    \Input{$\mathcal{H} \in \mathbb{R}^{n \times (T_{max}+1)}$: Profit matrix \\
    $T_{max} \in \mathbb{N}^+$: Total cache ways available}
    \Output{$(max\_profit, allocations)$: Optimal solution tuple}
    
    \BlankLine
    \textcolor{comment-green}{\tcp{Initialization phase}}
    $n \gets \text{rows}(\mathcal{H})$ \textcolor{comment-grey}{\tcp*{Number of L1 caches}}
    Initialize $\mathrm{dp}[0..n][0..T_{max}] \gets 0$ \;
    
    \For{$i \gets 1$ \KwTo $n$}{
        $\mathrm{dp}[i][0] \gets \sum_{k=0}^{i-1} \mathcal{H}[k][0]$ \textcolor{comment-grey}{\tcp*{Base: no allocation}}
}
    
    \BlankLine
    \textcolor{comment-green}{\tcp{DP table construction}}
    \For{$i \gets 1$ \KwTo $n$}{
        \For{$j \gets 1$ \KwTo $T_{max}$}{
            \textcolor{comment-grey}{\tcp{Default: no allocation to current cache}}
            $\mathrm{dp}[i][j] \gets \mathrm{dp}[i-1][j] + \mathcal{H}[i-1][0]$ \;
            
            \textcolor{comment-grey}{\tcp{Optimal allocation search}}
            \For{$k \gets 1$ \KwTo $j$}{
                \If{$\mathrm{dp}[i-1][j-k] + \mathcal{H}[i-1][k] > \mathrm{dp}[i][j]$}{
                    $\mathrm{dp}[i][j] \gets \mathrm{dp}[i-1][j-k] + \mathcal{H}[i-1][k]$
                }
            }
        }
    }
    
    \BlankLine
    \textcolor{comment-green}{\tcp{Backtrace for optimal allocations}}
    $allocations \gets [0]^n$ \textcolor{comment-grey}{\tcp*{Initialize allocation vector}}
    $j \gets T_{max}$ \;
    \For{$i \gets n$ \textbf{downto} $1$}{
        \For{$k \gets 0$ \KwTo $j$}{
            \If{$\mathrm{dp}[i][j] \equiv \mathrm{dp}[i-1][j-k] + \mathcal{H}[i-1][k]$}{
                $allocations[i-1] \gets k$ \;
                $j \gets j - k$ \;
                \textbf{break}\;
            }
        }
    }
    
    \Return $(\mathrm{dp}[n][T_{max}], allocations)$
}
\caption{Optimal Cache Way Allocation}
\label{alg:cache_allocation_algorithm}
\end{algorithm}
  \end{minipage}
  \begin{minipage}{0.33\linewidth}
    \centering
    \includegraphics[width=\linewidth]{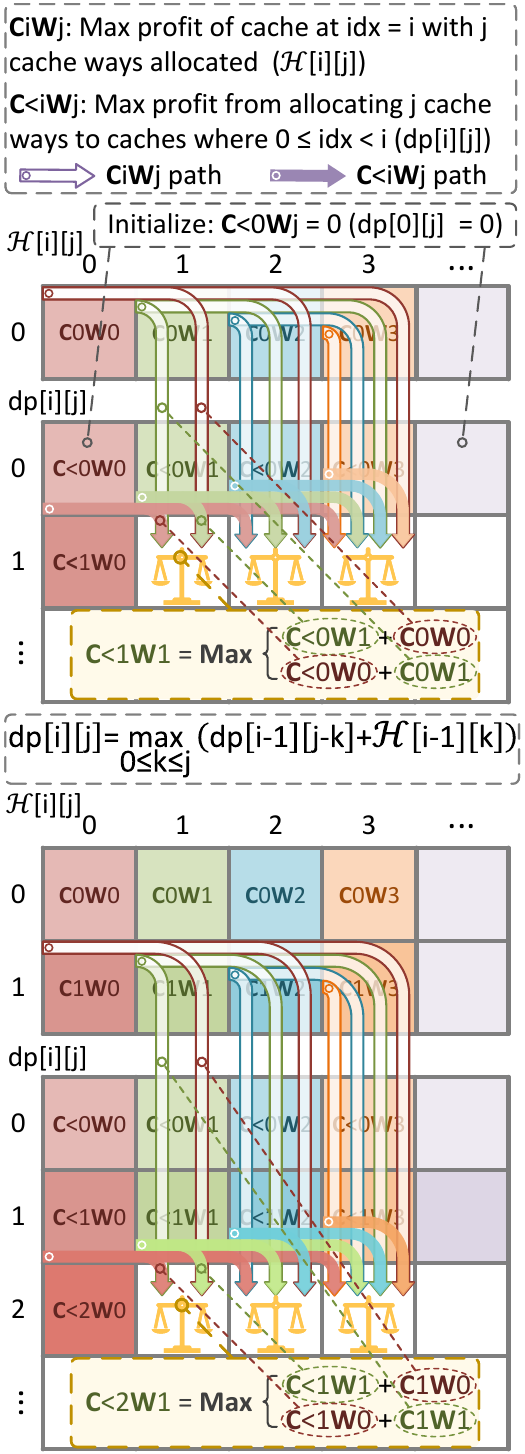}
    \caption{Illustration of DP Table Construction (cache count = 1 \& 2)}
    \label{fig:algorithm_fig}
  \end{minipage}
\end{figure*}

Using the obtained \( H_i(S_{i}) \), we solve the linear programming problem using Algorithm~\ref{alg:cache_allocation_algorithm} to find the optimal cache size allocation. Combining the corresponding cache line configurations yields the optimal cache configuration. Figure~\ref{fig:algorithm_fig} demonstrates dynamic cache way allocation during scaling by optimally distributing available ways between current/preceding caches under given states. 

\paragraph{Improvement: Redefining the Hit Rate}

The traditional cache hit rate is typically calculated as:

\[
\text{Hit Rate} = \frac{\text{Number of Hits}}{\text{Total Number of Accesses}}
\]

This formula calculates the average hit rate, abstracting details such as timing and access patterns. For example, in Figure~\ref{fig:memory_patterns}, the Regular and Irregular Access Mixed pattern (Figure~\hyperref[fig:memory_patterns]{\ref{fig:memory_patterns}g} and \hyperref[fig:memory_patterns]{\ref{fig:memory_patterns}h}) shows a significantly higher hit rate than the Irregular Access pattern (Figure~\hyperref[fig:memory_patterns]{\ref{fig:memory_patterns}d} and \hyperref[fig:memory_patterns]{\ref{fig:memory_patterns}e}). Yet, both patterns experience a similar number of misses and CGRA stalls within the same time frame. The hit rate of the Regular and Irregular Mixed pattern (Figure~\hyperref[fig:memory_patterns]{\ref{fig:memory_patterns}g} and \hyperref[fig:memory_patterns]{\ref{fig:memory_patterns}h}) is inflated by frequent regular accesses, leading to diminishing returns as cache size increases.

In our CGRA environment, memory accesses are synchronized across different caches, making it critical to avoid cache misses in all caches within a given time window. Relying on the traditional hit rate for cache allocation may lead to suboptimal cache size assignments, particularly under-allocating for caches with Regular and Irregular Access Mixed patterns (Figure~\hyperref[fig:memory_patterns]{\ref{fig:memory_patterns}g} and \hyperref[fig:memory_patterns]{\ref{fig:memory_patterns}h}), resulting in inefficient gains.

To address this, we redefine the cache miss rate as the \textbf{Time Miss Rate}:

\[
\text{Time Miss Rate} = \frac{\text{Number of Misses}}{\text{Length of Time Window}}
\]

And correspondingly, the \textbf{Time Hit Rate} is calculated as  1 - \text{Time Miss Rate}.

\section{Experimental Evaluation}
\label{sec:experimental_evaluation}

\begin{table*}[tbp]
\centering
\caption{Application kernels used in the evaluation}
\label{tab:benchmarks}
\resizebox{\linewidth}{!}{%
\begin{tabular}{|l|l|l|l|l|}
\hline
\textbf{Application} & \textbf{Kernel} & \textbf{Description} & \textbf{Input Data} & \textbf{Domain} \\ \hline
\multirow{4}{*}{Graph Convolutional Network} 
& \multirow{4}{*}{\texttt{aggregate}} 
& \multirow{4}{*}{Aggregates information from neighbors} 
& Citeseer~\cite{database} & \multirow{4}{*}{Graph Neural Networks} \\ 
& & & Cora~\cite{database} & \\
& & & PubMed~\cite{database} & \\
& & & OGBN-Arxiv~\cite{ogbn} & \\ \hline
OpenFOAM & \texttt{grad} & Gradient calculation and correction & \multirow{6}{*}{Random Data} & Computational Fluid Dynamics \\ \cline{1-3} \cline{5-5} 
Graclus & \texttt{perm\_sort} & Counting sort~\cite{sort} & & Graph Clustering \\ \cline{1-3} \cline{5-5} 
\multirow{2}{*}{MachSuite~\cite{MachSuite}} 
& \texttt{radix\_hist} & Radix sort (histogram) & & \multirow{2}{*}{Sorting Algorithms} \\
& \texttt{radix\_update} & Radix sort (update) & & \\ \cline{1-3} \cline{5-5} 
MiBench~\cite{MiBench} & \texttt{rgb} & Converts paletted color to RGB & & Image Processing \\ \cline{1-3} \cline{5-5} 
Berkeley Multimedia Workload~\cite{Berkeley} & \texttt{src2dest} & Audio processing & & Audio Processing \\ \hline
\end{tabular}%
}
\vspace{-10pt}
\end{table*}

In this section, we evaluate the effectiveness of the proposed methods through a series of experiments. \addr{We compare execution outputs both with and without our optimization technique to ensure the reliability of our approach.} We implement a cycle-accurate CGRA simulator based on the HyCUBE \delr{architecte}\addr{architecture}~\cite{HyCUBE}, integrating our redesigned memory subsystem and optimization techniques. Hardware employs varied cache configurations, each specified in respective sections. We select HyCUBE as our case study because it is a representative CGRA that effectively illustrates the challenges of this class of architectures, offers a relatively mature and comprehensive end-to-end compilation and simulation framework (albeit not originally cycle-accurate), and is attracting significant research interest.

Our benchmarks cover a diverse range of workloads, including GNN, Computational Fluid Dynamics, Sorting Algorithms, and Image Processing, as listed in Table~\ref{tab:benchmarks}. These benchmarks exhibit both compute-intensive characteristics and irregular memory access patterns, challenging traditional CGRA memory subsystems. They represent a variety of real-world applications where CGRAs can offer significant acceleration and allow us to thoroughly test the effectiveness of our runahead mechanism and cache reconfiguration techniques across diverse computational scenarios. 

We use a Graph Convolutional Network (GCN) model from PyTorch-Geometric~\cite{PyTorch}, focusing on the \texttt{aggregate} kernel. and select four widely used datasets—Citeseer, Cora, PubMed, and OGBN-Arxiv—with reduced dimensions to control simulation time. For other kernels, we generate randomized input data to ensure diverse and realistic memory access patterns.

While our benchmark covers a wide range of complex computational patterns, certain advanced operations, especially highly complex graph-based workloads, are limited by existing CGRA compilers. Enhancing these compilers is beyond the scope of this study. In future work, we plan to expand our benchmark suite to include more sophisticated workloads and further explore the potential of our techniques to address these challenges.

\subsection{Experimental Results and Analysis}
\label{sec:experimental_results_Analysis}

\begin{figure}[tbp]
    \centering
    \begin{subfigure}{0.75\linewidth}
        \centering
        \includegraphics[width=\linewidth]{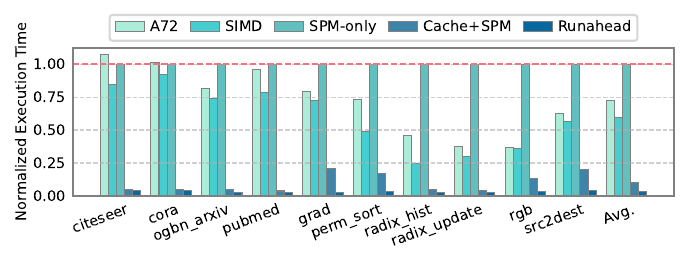}
        \vspace{-22pt}
        \caption{Normalized execution time for ARM Cortex-A72~\cite{armDocumentation}, SIMD~\cite{armDocumentation}, SPM-only~\cite{HyCUBE}, Cache-SPM and Runahead. Cache-SPM reduces SPM-related misses and memory bottlenecks, while runahead execution further improves performance.}
        \label{fig:execution_time_comparison}
    \end{subfigure}    
    \hfill
    \begin{subfigure}{0.24\linewidth}
        \centering
        \includegraphics[width=\linewidth]{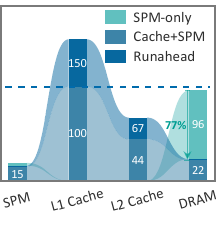}
        \vspace{-22pt}
        \caption{Normalized statistical access counts of memory components (L1-based).}
        \label{fig:bandwidth}
    \end{subfigure}
    \vspace{-20pt}
    \caption{Normalized Benchmark Execution Time and Memory Access Statistics}
    \vspace{-5pt}
    \label{fig:execution_time_comparison_bandwidth}
\end{figure}

Figure~\ref{fig:execution_time_comparison} compares the execution time of five systems: (1) ARM Cortex-\textbf{A72} (a high-efficiency core, with configurations detailed in Table~\ref{tab:arm_config})~\cite{armDocumentation}, (2) \textbf{SIMD}-accelerated A72 utilizing the Neon engine~\cite{armDocumentation}, (3) the original HyCUBE design—\textbf{SPM-only} (featuring a 133KB SPM)~\cite{HyCUBE}, (4) a \textbf{Cache+SPM} hybrid system \addr{with an overall storage capacity matching the SPM-only system (two 512-byte SPM modules, one 4 KB L1 cache, and one 128 KB L2 cache), as specified in Table~\ref{tab:config_cgra_cache} (base),}
and (5) a \textbf{Runahead}-enhanced Cache+SPM system.

\begin{table}[tbp]
\caption{A72 and SIMD configurations~\cite{armDocumentation}}
\label{tab:arm_config}
\scriptsize
\begin{tabular}{>{\bfseries}l c}
\toprule
Component & \multicolumn{1}{c}{\textbf{Specification}} \\
\midrule
\multirow{2}{*}{Core} 
& ARM Cortex-A72 (ARMv8-A) @ 1.8~GHz; Superscalar pipeline (variable-length, OoO);\\
& Hybrid BTB/GHB predictor; 128-bit NEON SIMD \\
L1 Cache & \textit{Instruction}: 48~KB/core (3-way); 
           \textit{Data}: 32~KB/core (2-way) \\
L2 Cache & 1~MB shared (16-way) \\
Memory & 8~GB LPDDR4-2400 SDRAM \\
\bottomrule
\end{tabular}
\end{table}

Prior research focuses on CGRA compilers and cores, leaving SPM as the default choice for data storage due to its simplicity, with limited attention to the memory subsystem. However, our evaluation reveals critical limitations in SPM-only implementations arising from asymmetric access characteristics: while SPM accesses incur near-zero latency, off-SPM memory operations suffer severe penalties. This dichotomy creates a performance degradation—as SPM miss rates increase with CGRA computation scaling, memory latency gradually dominates system behavior, reducing performance to memory-bound levels. Consequently, SPM-only implementations underperform even the A72 and SIMD systems in multiple kernels (Figure~\ref{fig:execution_time_comparison}), indicating that the CGRA's computational potential is bottlenecked by inefficient data supply from the memory subsystem.

In contrast, Cache+SPM shows superior scalability by using cache hierarchies. SPM misses are partially offset by cache hits. Benchmarks indicate that Cache+SPM achieves a 10× speedup over the \delr{area}\addr{size}-equivalent SPM-only design. It also outperforms A72 and SIMD across all kernels, with average speedups of 7.26× and 6.0×, respectively. The impact of cache configurations on system performance is analyzed in Section~\ref{sec:cache_config_impact}.

Despite its advantages, Cache+SPM remains susceptible to frequent CGRA stalls caused by irregular memory access patterns. Our runahead mechanism addresses this by proactively prefetching data during CGRA stalls triggered by cache misses. This approach achieves an average 3.04× speedup (up to 6.91×) over Cache+SPM, with detailed analysis in Section~\ref{sec:performance_runahead}.  

Figure~\ref{fig:bandwidth} quantifies memory access distribution across SPM-only, Cache+SPM, and Runahead systems. In SPM-only designs, frequent SPM misses force excessive DRAM (Dynamic Random Access Memory) accesses (high latency), resulting in low CGRA utilization due to prolonged stalls. Cache+SPM reduces DRAM accesses by 77\% by caching frequently reused data, significantly alleviating memory bottlenecks. Runahead optimizes by prefetching data during CGRA stall periods. Compared to SPM+Cache, overall memory subsystem access counts increase, but SPM and DRAM accesses rise less. This is because Runahead mainly activates on L1 cache misses, reducing SPM accesses. Additionally, Runahead's precise prefetching fetches required data from DRAM to cache in advance, limiting DRAM access increases.

In summary, Cache+SPM outperforms SPM-only designs by 10× with 77\% fewer DRAM accesses, while runahead prefetching delivers an additional 3.04× speedup (up to 6.91×). This highlights the need to integrate cache hierarchies and proactive prefetching to fully exploit CGRA's potential in compute-intensive and irregular memory access workloads.

\subsection{Impact of Cache Configuration on CGRA}
\label{sec:cache_config_impact}

\begin{figure*}[tbp]
    \centering
    \vspace{-10pt}
    \includegraphics[width=\linewidth]{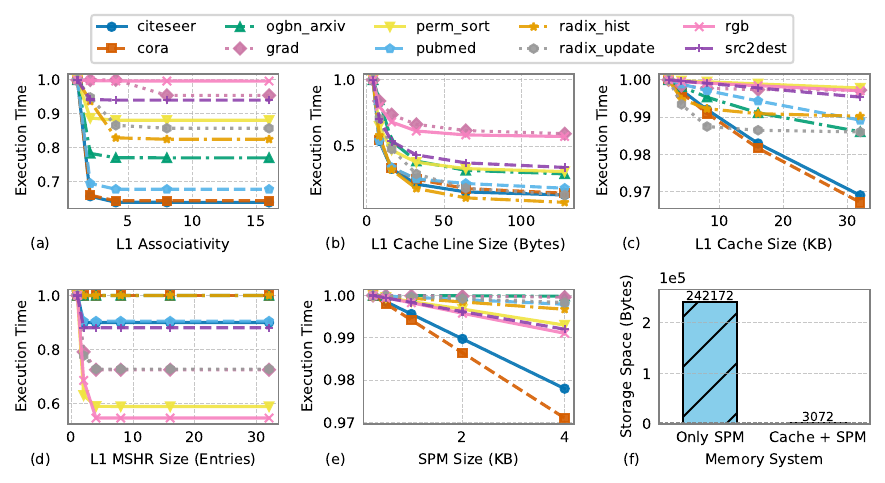}
    \caption{Execution Time is Normalized. The cache configuration significantly affects the execution time, with the Cache+SPM system consuming only 1.27\% of the storage space compared to the SPM-only system.}
    \vspace{-15pt}
    \label{fig:cache_config_effect}
\end{figure*}

Based on the configurations in Table~\ref{tab:config_cgra_cache} (Base), we analyze the effect of each parameter, noting that L1 and L2 cache line sizes adjust together. This experiment aims to establish well-calibrated parameters for a Cache+SPM system, providing a reliable basis for evaluating future optimizations.

Figure~\ref{fig:cache_config_effect} illustrates the effects of cache parameter configurations and SPM size on CGRA execution time, and compares the \delr{area}\addr{storage} requirements of SPM-only and Cache+SPM systems.

In Figure~\hyperref[fig:cache_config_effect]{\ref{fig:cache_config_effect}a}, the execution time reaches saturation when the L1 cache associativity is set to 8. In some tests, increasing associativity leads to notable performance improvements, primarily due to a reduction in cache thrashing.

Figure~\hyperref[fig:cache_config_effect]{\ref{fig:cache_config_effect}b} shows that the execution time saturates when the L1 cache line size reaches approximately 64 bytes. Although the CGRA processes irregular memory accesses during testing, a significant portion of memory accesses remains regular, such as accesses to \texttt{edge\_end} and \texttt{edge\_start} in the GCN \texttt{aggregate} kernel. These regular accesses are more sensitive to cache line size, and increasing the cache line size substantially improves the hit rate for these accesses.

In Figure~\hyperref[fig:cache_config_effect]{\ref{fig:cache_config_effect}c}, increasing the cache size allows more data to be stored for future use, particularly benefiting irregular memory accesses with good locality. However, increasing cache size does not always lead to significant performance improvements, as multiple PEs sharing a single L1 cache may cause frequent cache line evictions. Although regular memory accesses show less sensitivity to cache size, they have the potential to displace cache lines utilized by irregular accesses, which can result in performance deterioration.

The Miss Status Handling Register (MSHR) limits the number of cache misses that can be handled simultaneously. A CGRA can issue multiple memory requests in a single cycle. In a $4 \times 4$ HyCUBE architecture, where memory access occurs on the left PEs, up to four memory access requests can be generated per cycle. If all these requests miss in the cache and the MSHR entries are full, the system must wait for previous requests to complete before accepting new ones. As shown in Figure~\hyperref[fig:cache_config_effect]{\ref{fig:cache_config_effect}d}, when the MSHR size reaches 4, execution time saturates, as all cache miss requests can be processed concurrently, effectively hiding latency.

Figure~\hyperref[fig:cache_config_effect]{\ref{fig:cache_config_effect}e} shows that increasing SPM size has little impact on performance for kernels with large computational data. To analyze storage \delr{area }requirements, we compare SPM-only and Cache+SPM systems using the Cora dataset kernel. Figure~\hyperref[fig:cache_config_effect]{\ref{fig:cache_config_effect}f} reveals that Cache+SPM achieves similar performance to SPM-only while using only 1.27\% of the \delr{area}\addr{size}. This conclusion is drawn from a controlled experiment: a Cache+SPM configuration (2~KB L1 cache, 1~KB SPM, 64~B cache line, no L2 cache) is set up and measured. The SPM size in SPM-only is then scaled up until performance matches Cache+SPM. This highlights the \delr{area}\addr{memory} efficiency gap, emphasizing the limitations of SPM-only.

\begin{table}[tbp]
\centering
\caption{Hardware Configurations (Base vs. Cache+SPM/Runahead vs. Reconfig)~\cite{HyCUBE}}
\label{tab:config_cgra_cache}
\scriptsize
\begin{tabular}{>{\bfseries}l c c c}
\toprule
\textbf{Component/Parameter} & 
\textbf{Base} & 
\textbf{Cache+SPM/Runahead} & 
\textbf{Reconfig} \\
\midrule

CGRA Architecture & 4×4 HyCUBE @ 704~MHz & 4×4 HyCUBE @ 704~MHz & 8×8 HyCUBE @ 704~MHz \\
\cmidrule{1-4}

SPM Size & 2×512B & 2×512B & 4×2KB \\
\cmidrule{1-4}

\multicolumn{4}{l}{\textit{L1 Cache}} \\
Cache Size/Line Size & 4KB/32B & 4KB/64B & 4×4KB/64B \\
Associativity & 4-way & 4-way & 8-way \\
MSHR Entries & 16 & 16 & 4×16 \\
Hit Latency & \multicolumn{3}{c}{1 cycle} \\
\cmidrule{1-4}

\multicolumn{4}{l}{\textit{L2 Cache}} \\
Cache Size/Line Size & 128KB/32B & 128KB/64B & 128KB/128B \\
Associativity & \multicolumn{3}{c}{8-way} \\
Hit/Miss Latency & \multicolumn{3}{c}{8/80 cycles} \\
\bottomrule
\end{tabular}
\end{table}

\subsection{Performance Gains from Runahead Execution}
\label{sec:performance_runahead}

\begin{figure}[tbp]
    \centering
    \begin{minipage}{0.48\linewidth}
        \centering
        \includegraphics[width=\linewidth]{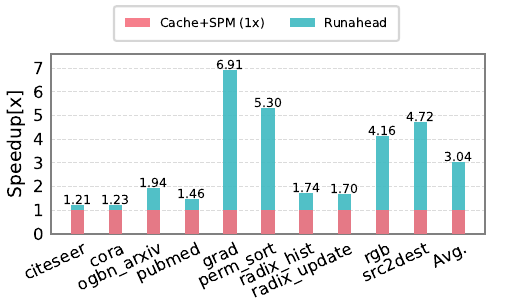}
        \vspace{-18pt}
        \caption{The runahead mechanism achieves to an average performance improvement of $3.04\times$ (up to $6.91\times$), with particularly significant gains in kernels exhibiting weaker locality.}
        \label{fig:runahead_perf}
    \end{minipage}
    \hfill
    \begin{minipage}{0.48\linewidth}
        \centering
        \includegraphics[width=\linewidth]{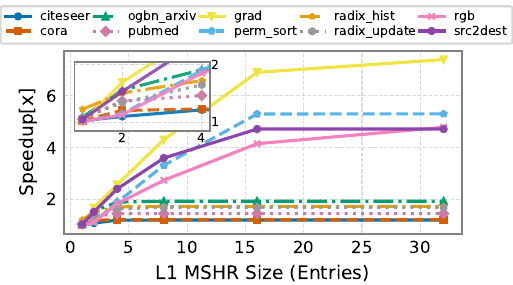}
        \caption{Kernels with weaker locality require more MSHRs for optimal performance. As the number of MSHRs increases, the benefits of the runahead mechanism gradually grow and saturate around 16.}
        \label{fig:mshr_impact}
    \end{minipage}
\end{figure}

Building on the findings from earlier experiments, we employ the relatively optimal Cache+SPM parameter configurations to rigorously evaluate the performance improvements brought by runahead execution. The detailed configurations are provided in Table~\ref{tab:config_cgra_cache} (Runahead), with a cache line size of 64~B to mitigate potential memory bandwidth pressure caused by larger cache lines.

As shown in Figure~\ref{fig:runahead_perf}, incorporating the runahead mechanism results in an average performance improvement of $3.04\times$ (up to $6.91\times$). Tests with a higher degree of randomness show more significant gains, as they experience more frequent cache misses, allowing runahead execution to substantially reduce the performance penalties associated with these misses, resulting in more significant and measurable performance improvements.

The effectiveness of runahead execution is closely linked to the size of the Miss Status Handling Register (MSHR). During runahead, a substantial number of cache requests are issued to facilitate prefetching, which relies on the MSHR to store pending cache miss requests. If the MSHR capacity is limited, it constrains the number of prefetches that can be effectively processed during runahead execution. To explore this relationship, we evaluate how the performance speedup from runahead execution varies with different MSHR sizes.

As illustrated in Figure~\ref{fig:mshr_impact}, the runahead speedup increases as the MSHR size grows, eventually saturating when the MSHR size reaches approximately 16. Tests with poorer memory locality tend to generate more cache misses, triggering runahead execution more frequently. During these periods, they issue a larger number of cache misses, requiring a greater number of MSHR entries, and as a result, achieve higher speedups from runahead execution. In contrast, tests with better memory locality trigger runahead execution less often, produce fewer cache misses during runahead, and thus require fewer MSHR entries. Consequently, the speedup from runahead execution in these tests is comparatively smaller.

\paragraph{Accuracy}
\label{sec:accuracy}

The CGRA runahead mechanism achieves precise prefetching through the meticulous tracking of dummy data, retaining only effective prefetch behaviors. This approach ensures that all prefetched data will be utilized in future computations, resulting in a prefetch accuracy of nearly 100\% for our CGRA runahead mechanism.

\begin{figure}[t]
    \centering
    \includegraphics[width=0.8\linewidth]{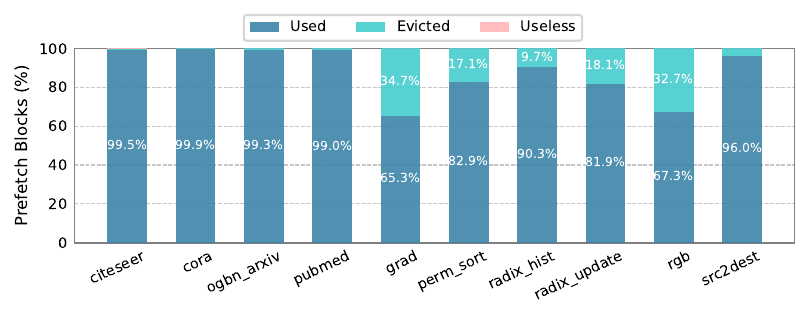}
    \vspace{-10pt}
    \caption{\addr{Distribution of prefetched cache blocks categorized as "Used" (useful blocks successfully utilized by the program), "Evicted" (useful blocks evicted before they can be used), and "Useless" (useless blocks). The near-zero proportion of "Useless" blocks indicates an almost 100\% prefetch accuracy. However, in highly random workloads (e.g., \texttt{grad}, \texttt{rgb}), the frequent prefetch and replacement of cache blocks leads to a higher incidence of evicted data, causing misses when the data is eventually needed.}}
    \vspace{-10pt}
    \label{fig:prefetch_distribution}
\end{figure}

\addr{Figure~\ref{fig:prefetch_distribution} illustrates the distribution of cache blocks prefetched during runahead. These cache blocks can be categorized into three groups based on whether they are needed by the program and whether they are successfully used, noting that some prefetched blocks may be evicted before they can be utilized.}

\addr{The majority of useless blocks (blocks not required by the program) arise under two circumstances. First, near the end of a computation, runahead execution may prefetch data that is never actually accessed. However, this end phase concludes quickly, so the number of unnecessary prefetched cache blocks remains negligible at a global scale. Second, if Read-After-Write (RAW) dependencies exist but the corresponding write operation cannot be completed during runahead due to missing data, any subsequent read from that address may fetch incorrect data, thereby driving execution into an invalid state. In such cases, the prefetched data could be useless. Nevertheless, because CGRAs employ a dataflow-based execution model, direct dependencies bypass main memory by traveling across the CGRA's interconnect network (between PEs), reducing memory accesses and boosting performance and efficiency. Consequently, most RAW dependencies are resolved through the interconnect rather than memory, thereby lowering the risk of incorrect states within our workloads. As a result, the proportion of useless cache blocks is nearly zero, which indirectly confirms that our prefetch strategy achieves close to 100\% accuracy.}

\addr{Nonetheless, we observe that some cache blocks needed by the program are evicted before use. This problem is more pronounced in tests characterized by higher randomness—such as \texttt{grad} and \texttt{rgb}—because frequent cache block prefetches and replacements increase the likelihood of prefetched data being evicted. To address this issue, adjusting relevant cache parameters—such as total size, cache line size, and associativity—to modify how data is mapped to the cache can help reduce conflicts and minimize the eviction of prefetched data.}

\paragraph{Coverage}
Figure~\ref{fig:runahead_coverage} illustrates the coverage, defined as the proportion of memory accesses addressed by the CGRA runahead mechanism. It also depicts the residual portion not mitigated by CGRA runahead prefetching, which manifests as demand misses in the CGRA. The average coverage achieved is 87\%. Kernels characterized by poor locality tend to exhibit a higher incidence of misses, presenting greater challenges in prefetching the corresponding data. Consequently, these kernels demonstrate relatively lower coverage rates. Notably, a higher coverage does not necessarily imply a greater speedup. This is because in the SPM+Cache system, the prefetching of runahead may overlap with the prefetching of the cache.

\subsection{Performance Gains from Cache Reconfiguration}

Based on prior experiments, we employ the relatively optimal cache parameter configurations for a rigorous evaluation of cache reconfiguration. The configurations are detailed in Table~\ref{tab:config_cgra_cache} (Reconfig).

We categorize the kernels into two groups based on the type of input data: real data and randomly generated data. Kernels using real data exhibit irregular memory accesses with some degree of locality, whereas those using randomly generated data generally lack locality. However, certain kernels, such as \texttt{radix\_hist} and \texttt{radix\_update}, involve operations like right-shift and bitwise AND on the input data, followed by memory accesses based on the computed results. This process imparts locality to these kernels.

\begin{figure}[t]
    \centering
    \begin{minipage}{0.48\linewidth}
        \centering
        \includegraphics[width=\linewidth]{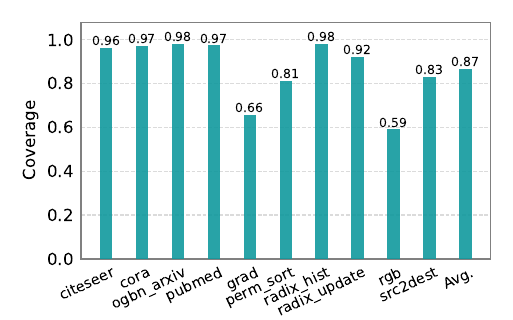}
        \caption{Percentage of memory accesses covered by CGRA runahead prefetching. Kernels with poor locality exhibit lower coverage due to the heightened difficulty of effective prefetching.}
        \label{fig:runahead_coverage}
        \vspace{-9pt}
    \end{minipage}
    \hfill
    \begin{minipage}{0.48\linewidth}
        \centering
        \includegraphics[width=\linewidth]{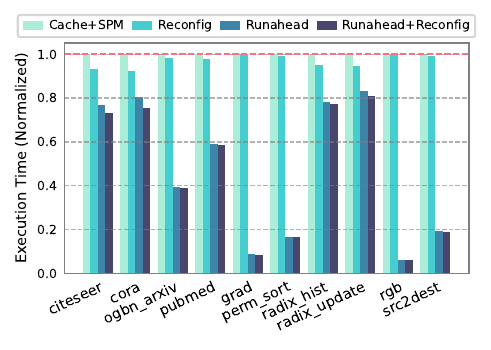}
        \vspace{-19pt}
        \caption{Cache reconfiguration consistently reduces runtime across various kernels, with those demonstrating higher memory locality achieving more significant performance improvements.}
        \vspace{9pt}
        \label{fig:cache_reconfig_gain}
        \vspace{-9pt}
    \end{minipage}
\end{figure}

As shown in Figure~\ref{fig:cache_reconfig_gain}, for kernels with real input data, cache reconfiguration reduces runtime by an average of 4.59\% (up to 7.79\%) without runahead, and by 3.22\% (up to 6.02\%) with runahead. For kernels with randomly generated data, cache reconfiguration reduces runtime by an average of 2.10\% (up to 5.26\%) without runahead, and by 1.58\% (up to 2.73\%) with runahead.

Since cache misses triggered by irregular memory accesses are the dominant performance bottleneck in CGRAs, our analysis focuses on how cache reconfiguration optimizes such accesses. Without runahead, cache enhancements offer limited benefits for kernels with poor locality, while they can provide more noticeable improvements for kernels with some locality. With runahead, increasing the cache size and adjusting the cache line size can mitigate the issue of prefetched data being prematurely evicted due to the large number of prefetches, leading to more substantial performance improvements for kernels with poor locality. For kernels with a degree of locality, similar cache modifications can also enhance performance. 

However, when irregular and regular memory accesses are interleaved, regular accesses may dominate and evict cache lines allocated for irregular accesses, diminishing the benefits of cache reconfiguration (as in \texttt{ogbn\_arxiv} and \texttt{pubmed}).

Based on our comprehensive analysis, we propose two optimization strategies for CGRA compilers. First, optimize memory access node mapping by avoiding the placement of regular and irregular memory accesses on the same PE or PEs sharing the same SPM to prevent interference. Second, optimize memory access timing by adjusting access schedules in shared-cache scenarios to enable interleaved cache utilization, thereby reducing performance degradation due to cache contention while maintaining overall throughput.

\subsection{Hardware Overhead Analysis}
\label{sec:hardware_overhead}

We employ a HyCUBE from the CGRA-ME project~\cite{CGRA_ME}, augmented with context save and restore logic. The HyCUBE supports 32-bit operands and up to 8 contexts. To assess the hardware overhead, we synthesize the modified design using Synopsys Design Compiler in a TSMC 28~nm process. The results indicate that our design incurs an area overhead of 14.78\% relative to the native HyCUBE. \addr{To avoid the potential influence of varying memory system sizes on our CGRA-optimization overhead analysis, this evaluation focuses solely on the CGRA itself, without involving the memory system.}

Since HyCUBE's hardware design is relatively simple, supporting only basic integer operations (e.g., \texttt{sub}, \texttt{add}, \texttt{multiply}, \texttt{and}, \texttt{or}, \texttt{xor}, \texttt{shl}) without division or modulus operations, implementing more complex features—such as floating-point arithmetic—would increase the area of the CGRA. In these scenarios, the relative overhead of our enhancements diminishes proportionally.

The overhead of cache reconfiguration is minimal, comprising two main components:
Cache Size Reconfiguration incurs a negligible overhead from additional permission registers assigned to each cache way(e.g., 4-bit permission registers for 32 cache ways add only 128 bits). 
Cache Line Size Reconfiguration involves managing both the physical cache line associated with a cache miss and other physical cache lines within the same virtual cache line. A counter is added to track processed physical cache lines, but reusing existing mechanisms keeps the overhead minimal.

Overall, the overhead of our design is well justified by the significant performance improvements, especially in handling irregular memory access patterns and processing large computational data efficiently, making the area-performance trade-off favorable.

\subsection{Area Breakdown}
\label{sec:area_breakdown}

\begin{figure}[tbp]
    \centering
    \begin{subfigure}{0.24\linewidth}
        \centering
        \includegraphics[width=\linewidth]{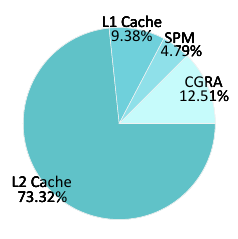}
        \caption{CGRA + Memory}
        \label{fig:pi_cgra_memory}
    \end{subfigure}    
    \hfill
    \begin{subfigure}{0.24\linewidth}
        \centering
        \includegraphics[width=\linewidth]{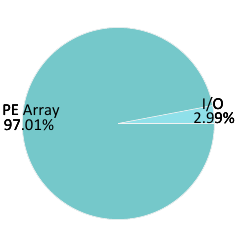}
        \caption{CGRA}
        \label{fig:pi_cgra}
    \end{subfigure}
    \hfill
    \begin{subfigure}{0.24\linewidth}
        \centering
        \includegraphics[width=\linewidth]{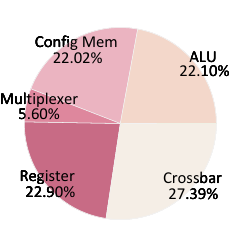}
        \caption{PE}
        \label{fig:pi_pe}
    \end{subfigure}
    \begin{subfigure}{0.24\linewidth}
        \centering
        \includegraphics[width=\linewidth]{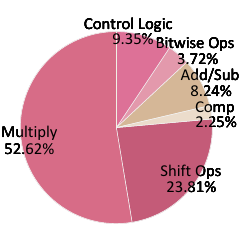}
        \caption{ALU}
        \label{fig:pi_alu}
    \end{subfigure}
    \caption{\addr{Area breakdown of the system configured according to Table~\ref{tab:config_cgra_cache} (Reconfig), including detailed breakdowns for the CGRA, a single PE within the PE array, and the ALU inside that PE.}}
    \label{fig:area_breakdown}
\end{figure}

\addr{Figure~\ref{fig:pi_cgra_memory} presents the area breakdown of the system configured as indicated in Table~\ref{tab:config_cgra_cache} (Reconfig). The L1 Cache occupies 9.38\% of the total area, while the 128 KB L2 Cache, sized to accommodate larger data, accounts for 73.32\%. Meanwhile, the 8×8 CGRA, featuring relatively simple logic, comprises 12.51\% of the area. Figure~\ref{fig:pi_cgra} details the CGRA's internal area distribution: the I/O circuitry (for CGRA configuration and memory transactions) occupies 2.99\%, while the PE array covers the remaining 97.01\%. Within this array, Figure~\ref{fig:pi_pe} shows the breakdown of a single PE. The crossbar, which supports inter-PE configurable interconnect, occupies 27.39\% of the PE area. The ALU accounts for 22.10\% of the PE area. Figure~\ref{fig:pi_alu} further details the ALU's composition: multiplication logic consumes 52.62\%, shift operations (\texttt{ashr}, \texttt{lshr}, \texttt{shl}) account for 23.81\%, and the control logic, which governs ALU operations, occupies 9.35\%. Bitwise operations (\texttt{and}, \texttt{or}, \texttt{xor}) and comparison require only minimal area.}

\section{Discussion}
\label{sec:discussion}

\addr{
In this section, we discuss various aspects of adaptability, scalability, and the CGRA compiler. While our experiments are based on HyCUBE, the proposed optimization techniques are not tied to any unique HyCUBE features but rather emerge from general CGRA design principles, offering strong adaptability. We further analyze the scalability of routing between the multi-cache system and the CGRA, as well as how our other optimization techniques affect the CGRA's area, power, and timing. Finally, we propose a potential path forward for CGRA compilers based on our findings.}

\subsection{Adaptability}
\label{sec:adaptability}

\addr{We select HyCUBE as our case study because it is a representative CGRA that effectively illustrates the challenges of this class of architectures and provides a relatively mature, end-to-end compilation and simulation framework. Notably, our proposed techniques do not depend on HyCUBE-specific features—such as multi-hop routing—but instead leverage core CGRA design principles~\cite{HyCUBE}, thereby demonstrating strong adaptability.}

\addr{The runahead execution mechanism is triggered when the CGRA stalls due to a cache miss, preserving the current state before runahead execution begins. This approach prefetches future data to reduce cache misses and subsequent stalls. Once runahead execution completes, the saved state is restored, ensuring CGRA correctness. As illustrated in Figure~\ref{fig:runahead_logic}, state preservation is achieved by duplicating the relevant register contents in backup registers. Upon entering runahead state, a dedicated runahead signal directs the CGRA to write the current state-related register data into these backup registers. After runahead execution, a multiplexer reconfigures the relevant registers' inputs to the backup register outputs, thereby recovering the saved state. To ensure precise prefetching, dummy data used during runahead are tracked by appending a bit to each datum, indicating whether it is dummy data. The ALU requires only a minor modification—a single OR gate—to propagate this bit from the input operands to the output. These additions operate in parallel with the original logic and remain decoupled, thereby minimizing disruptions to the existing design. Moreover, they do not rely on any HyCUBE-specific features but instead align with common CGRA design principles. Because CGRAs typically consist of multiple uniform PEs, this single modification can simply be replicated across all PEs, further streamlining the implementation.}

\addr{Since the cache reconfiguration technique operates as a separate memory subsystem optimization, it remains decoupled from the CGRA architecture and thus can be readily extended to other CGRAs. As a result, our proposed approaches are broadly applicable to a diverse range of CGRA designs.}

\subsection{Scalability}
\label{sec:scalability}

\addr{In this section, we analyze the proposed CGRA system's scalability with respect to four key dimensions: routing, area, power, and timing. Our results show that the design maintains linear complexity in each dimension, as described below.}

\addr{\textit{Routing Scalability.}
For the routing between the CGRA and the memory subsystem, as illustrated in Figure~\hyperref[fig:overview]{\ref{fig:overview}a~\textcolor{black}{\ding{187}}}, two PEs share a single cache and SPM pair via a crossbar, which obviates the need for direct connectivity between every PE and every cache/SPM block. Consequently, for an $n \times n$ CGRA, the routing complexity grows proportionally to $2n$.
Within the CGRA itself, the CGRA features a regular array structure in which each PE connects only to nearby PEs within a certain distance. Our optimization does not modify the underlying CGRA interconnect, thus preserving the original linear complexity as the CGRA scales.}

\addr{\textit{Area/power Scalability.}
Because the CGRA's regular, array-based architecture inherently scales linearly with the number of PEs, our modifications—which preserve this fundamental layout—maintain its linear area scalability. Furthermore, in array-based CGRA designs with consistent voltage and frequency, power consumption tends to increase proportionally to area; consequently, as the CGRA's area grows linearly, its power consumption follows a similar linear trend.}


\addr{\textit{Timing Scalability.}
The CGRA's critical path typically resides in the data path from a PE's output register to a successive PE's input register or within the PE's ALU logic~\cite{HyCUBE}. As discussed in Section~\ref{sec:adaptability} and illustrated in Figure~\ref{fig:runahead_logic}, the introduced enhancements (e.g., state save/restore logic, dummy data tracking) are logically simple and remain decoupled from the original CGRA data paths, thereby neither extending existing critical paths nor introducing new ones. Consequently, the critical path remains effectively unchanged as the CGRA scales.}

\subsection{CGRA Compiler}
\label{sec:cgra_compiler}

\addr{Although current CGRA compilers excel at mapping and scheduling well-structured loops efficiently, challenges remain—particularly in accommodating complex control flows, dynamic loops, irregular data structures, and memory‐aware scheduling. This section proposes several potential directions to advance CGRA compiler capabilities.}

\addr{A key avenue lies in leveraging MLIR (Multi-Level Intermediate Representation)~\cite{MLIR_scling}. 
Its comprehensive transformation passes readily handle intricate constructs,
while preserving domain‐specific abstractions. The dialect system also supports custom operations tailored for CGRAs, enabling the representation of diverse elements—from complex DFGs to various memory access patterns—under a single IR (Intermediate Representation)~\cite{MLIR_Moore}.}

\addr{Integrating LLMs (Large Language Models) into compiler toolchains marks another promising direction~\cite{llmcompiler,Autocomp}. 
Coupled with their understanding of code, this capability could inform transformation passes, suggest optimized rewrites, and refine pass ordering. By addressing challenges such as pointer aliasing and irregular loop constructs, 
LLM-guided optimizations show significant potential for tackling complex compilation scenarios.}

\addr{Equally pivotal is memory‐aware scheduling. Traditional methods often focus on the compute graph alone, overlooking memory contention. Incorporating explicit memory modeling into MLIR ensures that every pass accounts for memory hierarchies, from shared caches to specialized SPMs.}

\addr{Finally, unified abstractions that encompass computation and memory can bolster the synergy between scheduling and resource allocation. By representing both computational kernels and their memory requirements in a single IR, compilers can more effectively perform mapping, distinguishing predictable, regular operations from dynamic, irregular ones to mitigate cache conflicts.}

\addr{In summary, a holistic strategy integrating MLIR's flexible IR, LLM-guided optimizations, memory-aware scheduling, and unified abstractions of computation and memory charts a potential path forward for CGRA compiler evolution.}

\section{Related Work}
\label{sec:related_work}

Cheng Tan et al.~\cite{DRIPS} propose a CGRA architecture that dynamically balances its pipeline by adjusting resource allocation in response to variations in kernel execution times, thereby improving throughput in streaming applications. Their evaluation, focused on GCN applications using a small dataset that fits entirely within the SPM, primarily examines execution time variations due to input data changes. Consequently, their study does not address challenges related to SPM capacity limitations or issues arising from prefetching in kernels with irregular memory access patterns.

\addr{CASCADE~\cite{CASCADE} presents a decoupled access-execute CGRA design that separates address generation instructions (AGIs) from the primary computation instructions, thereby alleviating the on-array overhead and improving loop kernel performance. Its dedicated hardware, referred to as a "stream engine," handles the memory address generation for typical strided access patterns. 
However, while CASCADE handles regular strided accesses well, it can face challenges when dealing with irregular memory patterns that strain the stream engine's capabilities.
To address these scenarios, our work introduces a runahead-based prefetching scheme that efficiently mitigates irregular memory stalls by looking ahead to future accesses, regardless of their stride regularity. We supplement this approach with a reconfigurable cache subsystem, adjusting the caching strategy for different memory behavior profiles. By tailoring cache configurations per PE, our system accommodates a broader range of access patterns while preserving high computational efficiency. 
}

\textit{Runahead execution}~\cite{runahead1,runahead2,1183532,Vector_Runahead,Decoupled_Vector_Runahead} allows processors to continue executing future instructions when stalled due to long-latency memory accesses, enabling data prefetching and partially hiding memory latencies. While implementing runahead in processors is challenging due to their complex hardware designs and variable execution logic, CGRAs have simpler, more regular architectures by offloading computational complexity to the compiler. This simplicity facilitates the implementation of runahead in CGRAs, allowing them to more readily leverage its benefits.

Previous cache optimization efforts, like the self-adjusting cache architecture by Chuanjun Zhang et al.~\cite{self_tuning} and the \textit{Hopscotch} framework by Zhe Jiang et al.~\cite{Hopscotch}, focus on traditional CPU architectures. These studies optimize cache parameters or dynamically allocate cache capacity to improve performance and power consumption. However, CGRAs present unique challenges due to their fundamentally different microarchitectures. Implementing runahead execution in CGRAs requires not only non-blocking caches but also customized cache adaptations to match CGRA-specific characteristics—an aspect not typically addressed in traditional cache studies. In CPU architectures with interleaved regular access patterns, the inherent regularity is obscured, complicating efforts to leverage them for optimization. In contrast, CGRAs expose these distinct memory access patterns, unveiling novel and unique optimization opportunities. Our work addresses these challenges and harnesses new optimization opportunities by designing customized cache architectures and developing dynamic cache reconfiguration techniques. These techniques are tailored to better support CGRA runahead execution while also adapting to the unique memory access patterns of CGRA workloads, thereby enhancing cache utilization and significantly improving performance.

\section{Conclusion}

Modern workloads in domains such as Graph Neural Networks, high-performance computing, and databases feature numerous compute-intensive kernels well-suited for acceleration using CGRAs. However, many of these kernels exhibit irregular memory access patterns, rendering traditional CGRA systems based on SPM inadequate. Frequent memory misses cause CGRAs to become memory-bound, significantly limiting their broader adoption. To address this challenge, we design a memory subsystem for CGRAs that integrates cache mechanisms and implement a \textit{runahead execution} technique specifically tailored for CGRAs, resulting in substantial performance improvements. To ensure scalability, we introduce a \textit{multi-cache memory subsystem}, enabling the memory architecture to scale consistently with the CGRA. Furthermore, we observe that memory-accessing PEs in CGRAs exhibit unique memory access patterns, requiring personalized cache configurations. To meet these requirements and optimize cache utilization, we implement \textit{cache size reconfiguration} across L1 caches based on cache ways and \textit{cache line size reconfiguration} within each cache. Our comprehensive methodology addresses a critical bottleneck in CGRA design, dramatically expanding their computational scope to encompass a wider spectrum of applications, including those previously bounded by memory subsystem constraints. The proposed techniques not only augment CGRA functionality but also provide valuable heuristics for optimizing memory hierarchies in diverse accelerator architectures.

\bibliographystyle{ACM-Reference-Format}
\bibliography{references}


\begin{thebibliography}{46}


\ifx \showCODEN    \undefined \def \showCODEN     #1{\unskip}     \fi
\ifx \showISBNx    \undefined \def \showISBNx     #1{\unskip}     \fi
\ifx \showISBNxiii \undefined \def \showISBNxiii  #1{\unskip}     \fi
\ifx \showISSN     \undefined \def \showISSN      #1{\unskip}     \fi
\ifx \showLCCN     \undefined \def \showLCCN      #1{\unskip}     \fi
\ifx \shownote     \undefined \def \shownote      #1{#1}          \fi
\ifx \showarticletitle \undefined \def \showarticletitle #1{#1}   \fi
\ifx \showURL      \undefined \def \showURL       {\relax}        \fi
\providecommand\bibfield[2]{#2}
\providecommand\bibinfo[2]{#2}
\providecommand\natexlab[1]{#1}
\providecommand\showeprint[2][]{arXiv:#2}

\bibitem[{Arm Limited}(2025)]%
        {armDocumentation}
\bibfield{author}{\bibinfo{person}{{Arm Limited}}.} \bibinfo{year}{2025}\natexlab{}.
\newblock \bibinfo{title}{{Documentation – Arm Developer}}.
\newblock \bibinfo{howpublished}{\url{https://developer.arm.com/documentation/100095/0003/Introduction/Features}}.
\newblock
\newblock
\shownote{[Accessed 29-03-2025]}.


\bibitem[Bandara et~al\mbox{.}(2022)]%
        {Revamp}
\bibfield{author}{\bibinfo{person}{Thilini~Kaushalya Bandara}, \bibinfo{person}{Dhananjaya Wijerathne}, \bibinfo{person}{Tulika Mitra}, {and} \bibinfo{person}{Li-Shiuan Peh}.} \bibinfo{year}{2022}\natexlab{}.
\newblock \bibinfo{booktitle}{\emph{REVAMP: A Systematic Framework for Heterogeneous CGRA Realization}}.
\newblock
\urldef\tempurl%
\url{https://doi.org/10.5281/zenodo.5848404}
\showURL{%
\tempurl}


\bibitem[Blanas et~al\mbox{.}(2011)]%
        {hash}
\bibfield{author}{\bibinfo{person}{Spyros Blanas}, \bibinfo{person}{Yinan Li}, {and} \bibinfo{person}{Jignesh~M. Patel}.} \bibinfo{year}{2011}\natexlab{}.
\newblock \showarticletitle{Design and evaluation of main memory hash join algorithms for multi-core CPUs}. In \bibinfo{booktitle}{\emph{Proceedings of the 2011 ACM SIGMOD International Conference on Management of Data}} (Athens, Greece) \emph{(\bibinfo{series}{SIGMOD '11})}. \bibinfo{publisher}{Association for Computing Machinery}, \bibinfo{address}{New York, NY, USA}, \bibinfo{pages}{37–48}.
\newblock
\showISBNx{9781450306614}
\href{https://doi.org/10.1145/1989323.1989328}{doi:\nolinkurl{10.1145/1989323.1989328}}


\bibitem[Chin et~al\mbox{.}(2017)]%
        {CGRA_ME}
\bibfield{author}{\bibinfo{person}{S.~Alexander Chin}, \bibinfo{person}{Noriaki Sakamoto}, \bibinfo{person}{Allan Rui}, \bibinfo{person}{Jim Zhao}, \bibinfo{person}{Jin~Hee Kim}, \bibinfo{person}{Yuko Hara-Azumi}, {and} \bibinfo{person}{Jason Anderson}.} \bibinfo{year}{2017}\natexlab{}.
\newblock \showarticletitle{CGRA-ME: A unified framework for CGRA modelling and exploration}. In \bibinfo{booktitle}{\emph{2017 IEEE 28th International Conference on Application-specific Systems, Architectures and Processors (ASAP)}}. \bibinfo{pages}{184--189}.
\newblock
\href{https://doi.org/10.1109/ASAP.2017.7995277}{doi:\nolinkurl{10.1109/ASAP.2017.7995277}}


\bibitem[Cong et~al\mbox{.}(2011)]%
        {reuse}
\bibfield{author}{\bibinfo{person}{Jason Cong}, \bibinfo{person}{Hui Huang}, \bibinfo{person}{Chunyue Liu}, {and} \bibinfo{person}{Yi Zou}.} \bibinfo{year}{2011}\natexlab{}.
\newblock \showarticletitle{A reuse-aware prefetching scheme for scratchpad memory}. In \bibinfo{booktitle}{\emph{2011 48th ACM/EDAC/IEEE Design Automation Conference (DAC)}}. \bibinfo{pages}{960--965}.
\newblock


\bibitem[Cong et~al\mbox{.}(2014)]%
        {6861574}
\bibfield{author}{\bibinfo{person}{Jason Cong}, \bibinfo{person}{Hui Huang}, \bibinfo{person}{Chiyuan Ma}, \bibinfo{person}{Bingjun Xiao}, {and} \bibinfo{person}{Peipei Zhou}.} \bibinfo{year}{2014}\natexlab{}.
\newblock \showarticletitle{A Fully Pipelined and Dynamically Composable Architecture of CGRA}. In \bibinfo{booktitle}{\emph{2014 IEEE 22nd Annual International Symposium on Field-Programmable Custom Computing Machines}}. \bibinfo{pages}{9--16}.
\newblock
\href{https://doi.org/10.1109/FCCM.2014.12}{doi:\nolinkurl{10.1109/FCCM.2014.12}}


\bibitem[Cummins et~al\mbox{.}(2023)]%
        {llmcompiler}
\bibfield{author}{\bibinfo{person}{Chris Cummins}, \bibinfo{person}{Volker Seeker}, \bibinfo{person}{Dejan Grubisic}, \bibinfo{person}{Mostafa Elhoushi}, \bibinfo{person}{Youwei Liang}, \bibinfo{person}{Baptiste Roziere}, \bibinfo{person}{Jonas Gehring}, \bibinfo{person}{Fabian Gloeckle}, \bibinfo{person}{Kim Hazelwood}, \bibinfo{person}{Gabriel Synnaeve}, {and} \bibinfo{person}{Hugh Leather}.} \bibinfo{year}{2023}\natexlab{}.
\newblock \bibinfo{title}{Large Language Models for Compiler Optimization}.
\newblock
\showeprint[arxiv]{2309.07062}~[cs.PL]
\urldef\tempurl%
\url{https://arxiv.org/abs/2309.07062}
\showURL{%
\tempurl}


\bibitem[Dave and Shrivastava(2018)]%
        {Ccf}
\bibfield{author}{\bibinfo{person}{Shail Dave} {and} \bibinfo{person}{Aviral Shrivastava}.} \bibinfo{year}{2018}\natexlab{}.
\newblock \showarticletitle{Ccf: A cgra compilation framework}. In \bibinfo{booktitle}{\emph{Proc. 21st Design Autom. Test Europe (DATE)}}. \bibinfo{pages}{1}.
\newblock


\bibitem[Dundas and Mudge(1997)]%
        {runahead1}
\bibfield{author}{\bibinfo{person}{James Dundas} {and} \bibinfo{person}{Trevor Mudge}.} \bibinfo{year}{1997}\natexlab{}.
\newblock \showarticletitle{Improving data cache performance by pre-executing instructions under a cache miss}. In \bibinfo{booktitle}{\emph{Proceedings of the 11th International Conference on Supercomputing}} (Vienna, Austria) \emph{(\bibinfo{series}{ICS '97})}. \bibinfo{publisher}{Association for Computing Machinery}, \bibinfo{address}{New York, NY, USA}, \bibinfo{pages}{68–75}.
\newblock
\showISBNx{0897919025}
\href{https://doi.org/10.1145/263580.263597}{doi:\nolinkurl{10.1145/263580.263597}}


\bibitem[Fey and Lenssen(2019)]%
        {PyTorch}
\bibfield{author}{\bibinfo{person}{Matthias Fey} {and} \bibinfo{person}{Jan~Eric Lenssen}.} \bibinfo{year}{2019}\natexlab{}.
\newblock \showarticletitle{Fast graph representation learning with PyTorch Geometric}.
\newblock \bibinfo{journal}{\emph{arXiv preprint arXiv:1903.02428}} (\bibinfo{year}{2019}).
\newblock


\bibitem[Gobieski et~al\mbox{.}(2021)]%
        {Snafu}
\bibfield{author}{\bibinfo{person}{Graham Gobieski}, \bibinfo{person}{Ahmet~Oguz Atli}, \bibinfo{person}{Kenneth Mai}, \bibinfo{person}{Brandon Lucia}, {and} \bibinfo{person}{Nathan Beckmann}.} \bibinfo{year}{2021}\natexlab{}.
\newblock \showarticletitle{Snafu: an ultra-low-power, energy-minimal cgra-generation framework and architecture}. In \bibinfo{booktitle}{\emph{2021 ACM/IEEE 48th Annual International Symposium on Computer Architecture (ISCA)}}. IEEE, \bibinfo{pages}{1027--1040}.
\newblock


\bibitem[Guo and Luo(2020)]%
        {Pillars}
\bibfield{author}{\bibinfo{person}{Yijiang Guo} {and} \bibinfo{person}{Guojie Luo}.} \bibinfo{year}{2020}\natexlab{}.
\newblock \showarticletitle{Pillars: An integrated CGRA design framework}. In \bibinfo{booktitle}{\emph{Third Workshop on Open-Source EDA Technology (WOSET)}}. \bibinfo{pages}{1--5}.
\newblock


\bibitem[Guthaus et~al\mbox{.}(2001)]%
        {MiBench}
\bibfield{author}{\bibinfo{person}{M.R. Guthaus}, \bibinfo{person}{J.S. Ringenberg}, \bibinfo{person}{D. Ernst}, \bibinfo{person}{T.M. Austin}, \bibinfo{person}{T. Mudge}, {and} \bibinfo{person}{R.B. Brown}.} \bibinfo{year}{2001}\natexlab{}.
\newblock \showarticletitle{MiBench: A free, commercially representative embedded benchmark suite}. In \bibinfo{booktitle}{\emph{Proceedings of the Fourth Annual IEEE International Workshop on Workload Characterization. WWC-4 (Cat. No.01EX538)}}. \bibinfo{pages}{3--14}.
\newblock
\href{https://doi.org/10.1109/WWC.2001.990739}{doi:\nolinkurl{10.1109/WWC.2001.990739}}


\bibitem[Heroux and Dongarra(2013)]%
        {hpc}
\bibfield{author}{\bibinfo{person}{Michael~Allen Heroux} {and} \bibinfo{person}{Jack Dongarra}.} \bibinfo{year}{2013}\natexlab{}.
\newblock \bibinfo{booktitle}{\emph{Toward a new metric for ranking high performance computing systems.}}
\newblock \bibinfo{type}{{T}echnical {R}eport}. \bibinfo{institution}{Sandia National Lab.(SNL-NM), Albuquerque, NM (United States); University of~…}.
\newblock


\bibitem[Hong et~al\mbox{.}(2025)]%
        {Autocomp}
\bibfield{author}{\bibinfo{person}{Charles Hong}, \bibinfo{person}{Sahil Bhatia}, \bibinfo{person}{Alvin Cheung}, {and} \bibinfo{person}{Yakun~Sophia Shao}.} \bibinfo{year}{2025}\natexlab{}.
\newblock \bibinfo{title}{Autocomp: LLM-Driven Code Optimization for Tensor Accelerators}.
\newblock
\showeprint[arxiv]{2505.18574}~[cs.PL]
\urldef\tempurl%
\url{https://arxiv.org/abs/2505.18574}
\showURL{%
\tempurl}


\bibitem[Hu et~al\mbox{.}(2020)]%
        {ogbn}
\bibfield{author}{\bibinfo{person}{Weihua Hu}, \bibinfo{person}{Matthias Fey}, \bibinfo{person}{Marinka Zitnik}, \bibinfo{person}{Yuxiao Dong}, \bibinfo{person}{Hongyu Ren}, \bibinfo{person}{Bowen Liu}, \bibinfo{person}{Michele Catasta}, {and} \bibinfo{person}{Jure Leskovec}.} \bibinfo{year}{2020}\natexlab{}.
\newblock \showarticletitle{Open graph benchmark: datasets for machine learning on graphs}. In \bibinfo{booktitle}{\emph{Proceedings of the 34th International Conference on Neural Information Processing Systems}} (Vancouver, BC, Canada) \emph{(\bibinfo{series}{NIPS '20})}. \bibinfo{publisher}{Curran Associates Inc.}, \bibinfo{address}{Red Hook, NY, USA}, Article \bibinfo{articleno}{1855}, \bibinfo{numpages}{16}~pages.
\newblock
\showISBNx{9781713829546}


\bibitem[Jiang et~al\mbox{.}(2024)]%
        {Hopscotch}
\bibfield{author}{\bibinfo{person}{Zhe Jiang}, \bibinfo{person}{Kecheng Yang}, \bibinfo{person}{Nathan Fisher}, \bibinfo{person}{Nan Guan}, \bibinfo{person}{Neil~C. Audsley}, {and} \bibinfo{person}{Zheng Dong}.} \bibinfo{year}{2024}\natexlab{}.
\newblock \showarticletitle{Hopscotch: A Hardware-Software Co-Design for Efficient Cache Resizing on Multi-Core SoCs}.
\newblock \bibinfo{journal}{\emph{IEEE Transactions on Parallel and Distributed Systems}} \bibinfo{volume}{35}, \bibinfo{number}{1} (\bibinfo{year}{2024}), \bibinfo{pages}{89--104}.
\newblock
\href{https://doi.org/10.1109/TPDS.2023.3332711}{doi:\nolinkurl{10.1109/TPDS.2023.3332711}}


\bibitem[Karunaratne et~al\mbox{.}(2017)]%
        {HyCUBE}
\bibfield{author}{\bibinfo{person}{Manupa Karunaratne}, \bibinfo{person}{Aditi~Kulkarni Mohite}, \bibinfo{person}{Tulika Mitra}, {and} \bibinfo{person}{Li-Shiuan Peh}.} \bibinfo{year}{2017}\natexlab{}.
\newblock \showarticletitle{HyCUBE: A CGRA with reconfigurable single-cycle multi-hop interconnect}. In \bibinfo{booktitle}{\emph{2017 54th ACM/EDAC/IEEE Design Automation Conference (DAC)}}. \bibinfo{pages}{1--6}.
\newblock
\href{https://doi.org/10.1145/3061639.3062262}{doi:\nolinkurl{10.1145/3061639.3062262}}


\bibitem[Kim et~al\mbox{.}(2019)]%
        {Static}
\bibfield{author}{\bibinfo{person}{Youngbin Kim}, \bibinfo{person}{Kyoungwoo Lee}, {and} \bibinfo{person}{Aviral Shrivastava}.} \bibinfo{year}{2019}\natexlab{}.
\newblock \showarticletitle{Static Function Prefetching for Efficient Code Management on Scratchpad Memory}. In \bibinfo{booktitle}{\emph{2019 IEEE 37th International Conference on Computer Design (ICCD)}}. \bibinfo{pages}{350--358}.
\newblock
\href{https://doi.org/10.1109/ICCD46524.2019.00056}{doi:\nolinkurl{10.1109/ICCD46524.2019.00056}}


\bibitem[Kou et~al\mbox{.}(2020)]%
        {TAEM}
\bibfield{author}{\bibinfo{person}{Mingyang Kou}, \bibinfo{person}{Jiangyuan Gu}, \bibinfo{person}{Shaojun Wei}, \bibinfo{person}{Hailong Yao}, {and} \bibinfo{person}{Shouyi Yin}.} \bibinfo{year}{2020}\natexlab{}.
\newblock \showarticletitle{TAEM: Fast Transfer-Aware Effective Loop Mapping for Heterogeneous Resources on CGRA}. In \bibinfo{booktitle}{\emph{2020 57th ACM/IEEE Design Automation Conference (DAC)}}. \bibinfo{pages}{1--6}.
\newblock
\href{https://doi.org/10.1109/DAC18072.2020.9218668}{doi:\nolinkurl{10.1109/DAC18072.2020.9218668}}


\bibitem[Lattner et~al\mbox{.}(2020)]%
        {MLIR_Moore}
\bibfield{author}{\bibinfo{person}{Chris Lattner}, \bibinfo{person}{Mehdi Amini}, \bibinfo{person}{Uday Bondhugula}, \bibinfo{person}{Albert Cohen}, \bibinfo{person}{Andy Davis}, \bibinfo{person}{Jacques Pienaar}, \bibinfo{person}{River Riddle}, \bibinfo{person}{Tatiana Shpeisman}, \bibinfo{person}{Nicolas Vasilache}, {and} \bibinfo{person}{Oleksandr Zinenko}.} \bibinfo{year}{2020}\natexlab{}.
\newblock \bibinfo{title}{MLIR: A Compiler Infrastructure for the End of Moore's Law}.
\newblock
\showeprint[arxiv]{2002.11054}~[cs.PL]
\urldef\tempurl%
\url{https://arxiv.org/abs/2002.11054}
\showURL{%
\tempurl}


\bibitem[Lattner et~al\mbox{.}(2021)]%
        {MLIR_scling}
\bibfield{author}{\bibinfo{person}{Chris Lattner}, \bibinfo{person}{Mehdi Amini}, \bibinfo{person}{Uday Bondhugula}, \bibinfo{person}{Albert Cohen}, \bibinfo{person}{Andy Davis}, \bibinfo{person}{Jacques Pienaar}, \bibinfo{person}{River Riddle}, \bibinfo{person}{Tatiana Shpeisman}, \bibinfo{person}{Nicolas Vasilache}, {and} \bibinfo{person}{Oleksandr Zinenko}.} \bibinfo{year}{2021}\natexlab{}.
\newblock \showarticletitle{MLIR: Scaling Compiler Infrastructure for Domain Specific Computation}. In \bibinfo{booktitle}{\emph{2021 IEEE/ACM International Symposium on Code Generation and Optimization (CGO)}}. \bibinfo{pages}{2--14}.
\newblock
\href{https://doi.org/10.1109/CGO51591.2021.9370308}{doi:\nolinkurl{10.1109/CGO51591.2021.9370308}}


\bibitem[Lee et~al\mbox{.}(2015)]%
        {7167295}
\bibfield{author}{\bibinfo{person}{Hongsik Lee}, \bibinfo{person}{Dong Nguyen}, {and} \bibinfo{person}{Jongeun Lee}.} \bibinfo{year}{2015}\natexlab{}.
\newblock \showarticletitle{Optimizing stream program performance on CGRA-based systems?}. In \bibinfo{booktitle}{\emph{2015 52nd ACM/EDAC/IEEE Design Automation Conference (DAC)}}. \bibinfo{pages}{1--6}.
\newblock
\href{https://doi.org/10.1145/2744769.2744884}{doi:\nolinkurl{10.1145/2744769.2744884}}


\bibitem[Li et~al\mbox{.}(2022)]%
        {ChordMap}
\bibfield{author}{\bibinfo{person}{Zhaoying Li}, \bibinfo{person}{Dhananjaya Wijerathne}, \bibinfo{person}{Xianzhang Chen}, \bibinfo{person}{Anuj Pathania}, {and} \bibinfo{person}{Tulika Mitra}.} \bibinfo{year}{2022}\natexlab{}.
\newblock \showarticletitle{ChordMap: Automated Mapping of Streaming Applications Onto CGRA}.
\newblock \bibinfo{journal}{\emph{IEEE Transactions on Computer-Aided Design of Integrated Circuits and Systems}} \bibinfo{volume}{41}, \bibinfo{number}{2} (\bibinfo{year}{2022}), \bibinfo{pages}{306--319}.
\newblock
\href{https://doi.org/10.1109/TCAD.2021.3058313}{doi:\nolinkurl{10.1109/TCAD.2021.3058313}}


\bibitem[Liu et~al\mbox{.}(2024)]%
        {E2EMap}
\bibfield{author}{\bibinfo{person}{Dajiang Liu}, \bibinfo{person}{Yuxin Xia}, \bibinfo{person}{Jiaxing Shang}, \bibinfo{person}{Jiang Zhong}, \bibinfo{person}{Peng Ouyang}, {and} \bibinfo{person}{Shouyi Yin}.} \bibinfo{year}{2024}\natexlab{}.
\newblock \showarticletitle{E2EMap: End-to-End Reinforcement Learning for CGRA Compilation via Reverse Mapping}. In \bibinfo{booktitle}{\emph{2024 IEEE International Symposium on High-Performance Computer Architecture (HPCA)}}. \bibinfo{pages}{46--60}.
\newblock
\href{https://doi.org/10.1109/HPCA57654.2024.00015}{doi:\nolinkurl{10.1109/HPCA57654.2024.00015}}


\bibitem[Luo et~al\mbox{.}(2023)]%
        {ML_CGRA}
\bibfield{author}{\bibinfo{person}{Yixuan Luo}, \bibinfo{person}{Cheng Tan}, \bibinfo{person}{Nicolas~Bohm Agostini}, \bibinfo{person}{Ang Li}, \bibinfo{person}{Antonino Tumeo}, \bibinfo{person}{Nirav Dave}, {and} \bibinfo{person}{Tong Geng}.} \bibinfo{year}{2023}\natexlab{}.
\newblock \showarticletitle{ML-CGRA: An Integrated Compilation Framework to Enable Efficient Machine Learning Acceleration on CGRAs}. In \bibinfo{booktitle}{\emph{2023 60th ACM/IEEE Design Automation Conference (DAC)}}. \bibinfo{pages}{1--6}.
\newblock
\href{https://doi.org/10.1109/DAC56929.2023.10247873}{doi:\nolinkurl{10.1109/DAC56929.2023.10247873}}


\bibitem[Mei et~al\mbox{.}(2003)]%
        {ADRES}
\bibfield{author}{\bibinfo{person}{Bingfeng Mei}, \bibinfo{person}{Serge Vernalde}, \bibinfo{person}{Diederik Verkest}, \bibinfo{person}{Hugo~De Man}, {and} \bibinfo{person}{Rudy Lauwereins}.} \bibinfo{year}{2003}\natexlab{}.
\newblock \showarticletitle{ADRES: An Architecture with Tightly Coupled VLIW Processor and Coarse-Grained Reconfigurable Matrix}. In \bibinfo{booktitle}{\emph{International Conference on Field-Programmable Logic and Applications}}.
\newblock
\urldef\tempurl%
\url{https://api.semanticscholar.org/CorpusID:39182312}
\showURL{%
\tempurl}


\bibitem[Mutlu et~al\mbox{.}(2005)]%
        {runahead2}
\bibfield{author}{\bibinfo{person}{O. Mutlu}, \bibinfo{person}{Hyesoon Kim}, \bibinfo{person}{J. Stark}, {and} \bibinfo{person}{Y.N. Patt}.} \bibinfo{year}{2005}\natexlab{}.
\newblock \showarticletitle{On Reusing the Results of Pre-Executed Instructions in a Runahead Execution Processor}.
\newblock \bibinfo{journal}{\emph{IEEE Computer Architecture Letters}} \bibinfo{volume}{4}, \bibinfo{number}{1} (\bibinfo{year}{2005}), \bibinfo{pages}{2--2}.
\newblock
\href{https://doi.org/10.1109/L-CA.2005.1}{doi:\nolinkurl{10.1109/L-CA.2005.1}}


\bibitem[Mutlu et~al\mbox{.}(2003)]%
        {1183532}
\bibfield{author}{\bibinfo{person}{O. Mutlu}, \bibinfo{person}{J. Stark}, \bibinfo{person}{C. Wilkerson}, {and} \bibinfo{person}{Y.N. Patt}.} \bibinfo{year}{2003}\natexlab{}.
\newblock \showarticletitle{Runahead execution: an alternative to very large instruction windows for out-of-order processors}. In \bibinfo{booktitle}{\emph{The Ninth International Symposium on High-Performance Computer Architecture, 2003. HPCA-9 2003. Proceedings.}} \bibinfo{pages}{129--140}.
\newblock
\href{https://doi.org/10.1109/HPCA.2003.1183532}{doi:\nolinkurl{10.1109/HPCA.2003.1183532}}


\bibitem[Naithani et~al\mbox{.}(2021)]%
        {Vector_Runahead}
\bibfield{author}{\bibinfo{person}{Ajeya Naithani}, \bibinfo{person}{Sam Ainsworth}, \bibinfo{person}{Timothy~M. Jones}, {and} \bibinfo{person}{Lieven Eeckhout}.} \bibinfo{year}{2021}\natexlab{}.
\newblock \showarticletitle{Vector Runahead}. In \bibinfo{booktitle}{\emph{2021 ACM/IEEE 48th Annual International Symposium on Computer Architecture (ISCA)}}. \bibinfo{pages}{195--208}.
\newblock
\href{https://doi.org/10.1109/ISCA52012.2021.00024}{doi:\nolinkurl{10.1109/ISCA52012.2021.00024}}


\bibitem[Naithani et~al\mbox{.}(2023)]%
        {Decoupled_Vector_Runahead}
\bibfield{author}{\bibinfo{person}{Ajeya Naithani}, \bibinfo{person}{Jaime Roelandts}, \bibinfo{person}{Sam Ainsworth}, \bibinfo{person}{Timothy~M. Jones}, {and} \bibinfo{person}{Lieven Eeckhout}.} \bibinfo{year}{2023}\natexlab{}.
\newblock \showarticletitle{Decoupled Vector Runahead}. In \bibinfo{booktitle}{\emph{2023 56th IEEE/ACM International Symposium on Microarchitecture (MICRO)}}. \bibinfo{pages}{17--31}.
\newblock


\bibitem[Reagen et~al\mbox{.}(2014)]%
        {MachSuite}
\bibfield{author}{\bibinfo{person}{Brandon Reagen}, \bibinfo{person}{Robert Adolf}, \bibinfo{person}{Yakun~Sophia Shao}, \bibinfo{person}{Gu-Yeon Wei}, {and} \bibinfo{person}{David Brooks}.} \bibinfo{year}{2014}\natexlab{}.
\newblock \showarticletitle{MachSuite: Benchmarks for accelerator design and customized architectures}. In \bibinfo{booktitle}{\emph{2014 IEEE International Symposium on Workload Characterization (IISWC)}}. \bibinfo{pages}{110--119}.
\newblock
\href{https://doi.org/10.1109/IISWC.2014.6983050}{doi:\nolinkurl{10.1109/IISWC.2014.6983050}}


\bibitem[Salkhordeh et~al\mbox{.}(2018)]%
        {ReCA}
\bibfield{author}{\bibinfo{person}{Reza Salkhordeh}, \bibinfo{person}{Shahriar Ebrahimi}, {and} \bibinfo{person}{Hossein Asadi}.} \bibinfo{year}{2018}\natexlab{}.
\newblock \showarticletitle{ReCA: An Efficient Reconfigurable Cache Architecture for Storage Systems with Online Workload Characterization}.
\newblock \bibinfo{journal}{\emph{IEEE Transactions on Parallel and Distributed Systems}} \bibinfo{volume}{29}, \bibinfo{number}{7} (\bibinfo{year}{2018}), \bibinfo{pages}{1605--1620}.
\newblock
\href{https://doi.org/10.1109/TPDS.2018.2796100}{doi:\nolinkurl{10.1109/TPDS.2018.2796100}}


\bibitem[Sen et~al\mbox{.}(2008)]%
        {database}
\bibfield{author}{\bibinfo{person}{Prithviraj Sen}, \bibinfo{person}{Galileo Namata}, \bibinfo{person}{Mustafa Bilgic}, \bibinfo{person}{Lise Getoor}, \bibinfo{person}{Brian Gallagher}, {and} \bibinfo{person}{Tina Eliassi-Rad}.} \bibinfo{year}{2008}\natexlab{}.
\newblock \showarticletitle{Collective Classification in Network Data}. In \bibinfo{booktitle}{\emph{The AI Magazine}}.
\newblock
\urldef\tempurl%
\url{https://api.semanticscholar.org/CorpusID:62016134}
\showURL{%
\tempurl}


\bibitem[Seward(1954)]%
        {sort}
\bibfield{author}{\bibinfo{person}{Harold~Herbert Seward}.} \bibinfo{year}{1954}\natexlab{}.
\newblock \emph{\bibinfo{title}{Information sorting in the application of electronic digital computers to business operations}}.
\newblock \bibinfo{thesistype}{Ph.\,D. Dissertation}. \bibinfo{school}{Massachusetts Institute of Technology. Department of Electrical Engineering}.
\newblock


\bibitem[Singh et~al\mbox{.}(2000)]%
        {MorphoSys}
\bibfield{author}{\bibinfo{person}{Hartej Singh}, \bibinfo{person}{Ming-Hau Lee}, \bibinfo{person}{Guangming Lu}, \bibinfo{person}{Nader Bagherzadeh}, \bibinfo{person}{Fadi~J. Kurdahi}, {and} \bibinfo{person}{Eliseu M.~Chaves Filho}.} \bibinfo{year}{2000}\natexlab{}.
\newblock \showarticletitle{MorphoSys: An Integrated Reconfigurable System for Data-Parallel and Computation-Intensive Applications}.
\newblock \bibinfo{journal}{\emph{IEEE Trans. Comput.}} \bibinfo{volume}{49}, \bibinfo{number}{5} (\bibinfo{date}{may} \bibinfo{year}{2000}), \bibinfo{pages}{465–481}.
\newblock
\showISSN{0018-9340}
\href{https://doi.org/10.1109/12.859540}{doi:\nolinkurl{10.1109/12.859540}}


\bibitem[Slingerland and Smith(2002)]%
        {Berkeley}
\bibfield{author}{\bibinfo{person}{Nathan~T Slingerland} {and} \bibinfo{person}{Alan~Jay Smith}.} \bibinfo{year}{2002}\natexlab{}.
\newblock \showarticletitle{Design and characterization of the Berkeley multimedia workload}.
\newblock \bibinfo{journal}{\emph{Multimedia Systems}} \bibinfo{volume}{8}, \bibinfo{number}{4} (\bibinfo{year}{2002}), \bibinfo{pages}{315--327}.
\newblock


\bibitem[Tan et~al\mbox{.}(2022)]%
        {DRIPS}
\bibfield{author}{\bibinfo{person}{Cheng Tan}, \bibinfo{person}{Nicolas~Bohm Agostini}, \bibinfo{person}{Tong Geng}, \bibinfo{person}{Chenhao Xie}, \bibinfo{person}{Jiajia Li}, \bibinfo{person}{Ang Li}, \bibinfo{person}{Kevin~J. Barker}, {and} \bibinfo{person}{Antonino Tumeo}.} \bibinfo{year}{2022}\natexlab{}.
\newblock \showarticletitle{DRIPS: Dynamic Rebalancing of Pipelined Streaming Applications on CGRAs}. In \bibinfo{booktitle}{\emph{2022 IEEE International Symposium on High-Performance Computer Architecture (HPCA)}}. \bibinfo{pages}{304--316}.
\newblock
\href{https://doi.org/10.1109/HPCA53966.2022.00030}{doi:\nolinkurl{10.1109/HPCA53966.2022.00030}}


\bibitem[Tan et~al\mbox{.}(2020)]%
        {OpenCGRA}
\bibfield{author}{\bibinfo{person}{Cheng Tan}, \bibinfo{person}{Chenhao Xie}, \bibinfo{person}{Ang Li}, \bibinfo{person}{Kevin~J. Barker}, {and} \bibinfo{person}{Antonino Tumeo}.} \bibinfo{year}{2020}\natexlab{}.
\newblock \showarticletitle{OpenCGRA: An Open-Source Unified Framework for Modeling, Testing, and Evaluating CGRAs}. In \bibinfo{booktitle}{\emph{2020 IEEE 38th International Conference on Computer Design (ICCD)}}. \bibinfo{pages}{381--388}.
\newblock
\href{https://doi.org/10.1109/ICCD50377.2020.00070}{doi:\nolinkurl{10.1109/ICCD50377.2020.00070}}


\bibitem[Torng et~al\mbox{.}(2021)]%
        {9407079}
\bibfield{author}{\bibinfo{person}{Christopher Torng}, \bibinfo{person}{Peitian Pan}, \bibinfo{person}{Yanghui Ou}, \bibinfo{person}{Cheng Tan}, {and} \bibinfo{person}{Christopher Batten}.} \bibinfo{year}{2021}\natexlab{}.
\newblock \showarticletitle{Ultra-Elastic CGRAs for Irregular Loop Specialization}. In \bibinfo{booktitle}{\emph{2021 IEEE International Symposium on High-Performance Computer Architecture (HPCA)}}. \bibinfo{pages}{412--425}.
\newblock
\href{https://doi.org/10.1109/HPCA51647.2021.00042}{doi:\nolinkurl{10.1109/HPCA51647.2021.00042}}


\bibitem[Wijerathne et~al\mbox{.}(2019)]%
        {CASCADE}
\bibfield{author}{\bibinfo{person}{Dhananjaya Wijerathne}, \bibinfo{person}{Zhaoying Li}, \bibinfo{person}{Manupa Karunarathne}, \bibinfo{person}{Anuj Pathania}, {and} \bibinfo{person}{Tulika Mitra}.} \bibinfo{year}{2019}\natexlab{}.
\newblock \showarticletitle{CASCADE: High Throughput Data Streaming via Decoupled Access-Execute CGRA}.
\newblock \bibinfo{journal}{\emph{ACM Trans. Embed. Comput. Syst.}} \bibinfo{volume}{18}, \bibinfo{number}{5s}, Article \bibinfo{articleno}{50} (\bibinfo{date}{Oct.} \bibinfo{year}{2019}), \bibinfo{numpages}{26}~pages.
\newblock
\showISSN{1539-9087}
\href{https://doi.org/10.1145/3358177}{doi:\nolinkurl{10.1145/3358177}}


\bibitem[Wijerathne et~al\mbox{.}(2022)]%
        {Morpher}
\bibfield{author}{\bibinfo{person}{Dhananjaya Wijerathne}, \bibinfo{person}{Zhaoying Li}, \bibinfo{person}{Manupa Karunaratne}, \bibinfo{person}{Li-Shiuan Peh}, {and} \bibinfo{person}{Tulika Mitra}.} \bibinfo{year}{2022}\natexlab{}.
\newblock \showarticletitle{Morpher: An open-source integrated compilation and simulation framework for cgra}. In \bibinfo{booktitle}{\emph{Fifth Workshop on Open-Source EDA Technology (WOSET)}}.
\newblock


\bibitem[Woodruff et~al\mbox{.}(2023)]%
        {Rewriting_History}
\bibfield{author}{\bibinfo{person}{Jackson Woodruff}, \bibinfo{person}{Thomas Koehler}, \bibinfo{person}{Alexander Brauckmann}, \bibinfo{person}{Chris Cummins}, \bibinfo{person}{Sam Ainsworth}, {and} \bibinfo{person}{Michael F.~P. O'Boyle}.} \bibinfo{year}{2023}\natexlab{}.
\newblock \showarticletitle{Rewriting History: Repurposing Domain-Specific CGRAs}.
\newblock \bibinfo{journal}{\emph{CoRR}}  \bibinfo{volume}{abs/2309.09112} (\bibinfo{year}{2023}).
\newblock
\href{https://doi.org/10.48550/ARXIV.2309.09112}{doi:\nolinkurl{10.48550/ARXIV.2309.09112}}
\showeprint[arXiv]{2309.09112}


\bibitem[Zhang et~al\mbox{.}(2004)]%
        {self_tuning}
\bibfield{author}{\bibinfo{person}{Chuanjun Zhang}, \bibinfo{person}{Frank Vahid}, {and} \bibinfo{person}{Roman Lysecky}.} \bibinfo{year}{2004}\natexlab{}.
\newblock \showarticletitle{A self-tuning cache architecture for embedded systems}.
\newblock \bibinfo{journal}{\emph{ACM Trans. Embed. Comput. Syst.}} \bibinfo{volume}{3}, \bibinfo{number}{2} (\bibinfo{date}{may} \bibinfo{year}{2004}), \bibinfo{pages}{407–425}.
\newblock
\showISSN{1539-9087}
\href{https://doi.org/10.1145/993396.993405}{doi:\nolinkurl{10.1145/993396.993405}}


\bibitem[Zhao et~al\mbox{.}(2020)]%
        {9075353}
\bibfield{author}{\bibinfo{person}{Zhongyuan Zhao}, \bibinfo{person}{Weiguang Sheng}, \bibinfo{person}{Qin Wang}, \bibinfo{person}{Wenzhi Yin}, \bibinfo{person}{Pengfei Ye}, \bibinfo{person}{Jinchao Li}, {and} \bibinfo{person}{Zhigang Mao}.} \bibinfo{year}{2020}\natexlab{}.
\newblock \showarticletitle{Towards Higher Performance and Robust Compilation for CGRA Modulo Scheduling}.
\newblock \bibinfo{journal}{\emph{IEEE Transactions on Parallel and Distributed Systems}} \bibinfo{volume}{31}, \bibinfo{number}{9} (\bibinfo{year}{2020}), \bibinfo{pages}{2201--2219}.
\newblock
\href{https://doi.org/10.1109/TPDS.2020.2989149}{doi:\nolinkurl{10.1109/TPDS.2020.2989149}}


\bibitem[Zhou et~al\mbox{.}(2020)]%
        {gnn}
\bibfield{author}{\bibinfo{person}{Jie Zhou}, \bibinfo{person}{Ganqu Cui}, \bibinfo{person}{Shengding Hu}, \bibinfo{person}{Zhengyan Zhang}, \bibinfo{person}{Cheng Yang}, \bibinfo{person}{Zhiyuan Liu}, \bibinfo{person}{Lifeng Wang}, \bibinfo{person}{Changcheng Li}, {and} \bibinfo{person}{Maosong Sun}.} \bibinfo{year}{2020}\natexlab{}.
\newblock \showarticletitle{Graph neural networks: A review of methods and applications}.
\newblock \bibinfo{journal}{\emph{AI Open}}  \bibinfo{volume}{1} (\bibinfo{year}{2020}), \bibinfo{pages}{57--81}.
\newblock
\showISSN{2666-6510}
\href{https://doi.org/10.1016/j.aiopen.2021.01.001}{doi:\nolinkurl{10.1016/j.aiopen.2021.01.001}}


\end{thebibliography}

\end{document}